\documentclass[fleqn,usenatbib]{mnras}

\usepackage{newtxtext,newtxmath}

\usepackage[T1]{fontenc}
\usepackage{ae,aecompl}
\usepackage{xcolor}

\usepackage{graphicx}	
\usepackage{amsmath}	
\usepackage{amssymb}	
\usepackage{subfigure}

\title[Supergranular turbulence in a quiet Sun]{Supergranular turbulence in a quiet Sun: Lagrangian coherent structures}

\author[A. C.-L. Chian, et al.]{
Abraham C.-L. Chian $^{1,2,3,4}$,\thanks{E-mail: abraham.chian@gmail.com}
Suzana S. A. Silva $^{4}$,
Erico L. Rempel $^{3,4}$,
\newauthor
Milan Go{\v{s}}i{\'{c}}$^{5,6}$,
Luis R. Bellot Rubio$^{7}$,
Kanya Kusano$^{2}$,
\newauthor
Rodrigo A. Miranda$^{8,9}$,
and Iker S. Requerey$^{10}$
\\
$^{1}$School of Mathematical Sciences, University of Adelaide, Adelaide, SA 5005, Australia\\
$^{2}$Institute for Space-Earth Environmental Research, Nagoya University, Furo-cho, Chikusa-ku, Nagoya 464-8601, Japan\\
$^{3}$National Institute for Space Research (INPE), PO Box 515, S\~{a}o Jos\'e dos Campos, SP 12227-010, Brazil\\
$^{4}$ Institute of Aeronautical Technology (ITA), World Institute for Space Environment Research (WISER), S\~{a}o Jos\'{e} dos Campos,\\ 
SP 12228-900, Brazil\\
$^{5}$ Lockheed Martin Solar and Astrophysics Laboratory, Palo Alto, CA 94304, USA\\
$^{6}$ Bay Area Environmental Research Institute, Moffett Field, CA 94035, USA\\
$^{7}$ Instituto de Astrof\'{i}sica de Andaluc\'{i}a (CSIC), Apdo. de Correos 3004, 18080 Granada, Spain\\
$^{8}$ UnB-Gama Campus, University of Bras\'{i}lia (UnB), Bras\'{i}lia DF 70910-900, Brazil\\
$^{9}$ Plasma Physics Laboratory, Institute of Physics, University of Bras\'{i}lia (UnB), Bras\'{i}lia DF 70910-900, Brazil\\
$^{10}$ Max Planck Institute for Solar System Research, Justus-von-Liebig-Weg 3, D-37077 G\"ottingen, Germany
}
\date{Accepted XXX. Received YYY; in original form ZZZ}

\pubyear{2019}

\begin{document}
\label{firstpage}
\pagerange{\pageref{firstpage}--\pageref{lastpage}}
\maketitle
\begin{abstract}
The quiet Sun exhibits a wealth of magnetic activities that are fundamental for our understanding of solar and astrophysical magnetism. The magnetic fields in the quiet Sun are observed to evolve coherently, interacting with each other to form distinguished structures as they are advected by the horizontal photospheric flows. We study coherent structures in photospheric flows in a region of quiet Sun consisted of supergranules. Supergranular turbulence is investigated by detecting hyperbolic and elliptic Lagrangian coherent structures (LCS) using the horizontal velocity fields derived from Hinode
intensity maps. Repelling/attracting LCS are found by computing the forward/backward finite-time Lyapunov exponent (FTLE). The Lagrangian centre of a supergranular cell is given by the local maximum of the forward FTLE; the Lagrangian boundaries of supergranular cells are given by the ridges of the backward FTLE. Objective velocity vortices are found by calculating the Lagrangian-averaged vorticity deviation, and false vortices are filtered by applying a criterion given by the displacement vector. The Lagrangian centres of neighboring supergranular cells are interconnected by ridges of the repelling LCS, which provide the transport barriers that allow the formation of vortices and the concentration of strong magnetic fields in the valleys of the repelling LCS. The repelling LCS also reveal the most likely sites for magnetic reconnection. We show that the ridges of the attracting LCS expose the locations of the sinks of photospheric flows at supergranular junctions, which are the preferential sites for the formation of kG magnetic flux tubes and persistent vortices. 
\end{abstract}




\section{Introduction}
The first report of a solar flare was published by \cite{Carrington1859} based on his observation on September 1, 1859 of the outburst of two patches of intensely bright and white light in an active region involving a group of sunspots at the Sun's disk. Later, \cite{Parker1955} showed that a magnetic flux tube driven by dynamo in the convective zone under the Sun's surface is buoyant and can rise to the photosphere, leading to the formation of sunspots. \cite{Parker1957} and \cite{Gold1960} suggested that solar flares can originate from the dissipation of energy associated with electric currents when two magnetic flux tubes reconnect in the solar atmosphere. Recent observations and numerical simulations have confirmed that coherent structures such as magnetic flux tubes (or ropes) and velocity vortices play a key role in the genesis and evolution of sunspots as well as in magnetic reconnections in solar corona,  responsible for the production of solar flares and coronal mass ejections \citep{Kamide2007}.

The origin of solar cycles, and therefore activity, lies below the photosphere where plasma motions associated with turbulent convection \citep{Hotta2019} produce magnetic fields. The photosphere is permeated by convection cells, called granules, with a typical spatial scale of  $\sim$1 Mm and a lifetime of a few minutes. On scales larger than granules, there is supergranulation, which is a complex nonlinear dynamical phenomenon taking place in the solar photosphere that exhibits a cellular flow pattern with a typical horizontal scale of  $\sim$30-35 Mm, a lifetime of  $\sim$24-48 hr, a strong horizontal flow velocity of $\sim$300-400 m/s and a weaker vertical velocity of  $\sim$20-30 m/s \citep{Rincon2018}. The origin of supergranulation can be attributed to thermal magnetoconvection such as magnetised Raleigh-B\'enard convection or large-scale instabilities, inverse cascades and collective interactions. 

Solar photosphere consists of active and quiet regions. Active regions refer to areas of strong kilogauss magnetic fields composed of sunspots, pores and plages. In addition, we also recognise the so-called quiet Sun regions which are composed of network (NE) and internetwork (IN) areas \citep{Bellot2019}. Photospheric NE outlines the supergranular cell boundaries, whereas the IN represents the interior of supergranular cells. The NE has strong fields of $\sim$1 kG \citep{Stenflo1973}, whereas the IN has weaker fields of $\sim$1 hG \citep{Orozco2007a, Orozco2007b}. The IN consists of mixed-polarity magnetic elements which are advected by supergranular flows from the inner parts of a supergranular cell toward its boundaries \citep{Orozco2012}, thus supplying the total NE flux in $\sim$10 h \citep{Gosic2014}. NE magnetic elements, with a characteristic size of $\sim$100-300 km and a downward velocity of $\sim$ 0.5 km/s, consist of vertical kG flux tubes that expand with height \citep{Stenflo1973}. They are very persistent and may last up to dozens of hours until they are diffused either by flux dispersal or cancellation with nearby elements \citep{Liu1994}. The persistence of a NE magnetic flux tube may be related to the vortical flow surrounding the magnetic structures. Such flows are able to stabilize the flux tube by the centrifugal force of its whirly motion \citep{Schussler1984}, as evidenced by the event found by \cite{Attie2009},  where a NE magnetic element appears to be co-aligned with a velocity vortex and the first observation of correlated magnetic vortices and photospheric downflows by \cite{Requerey2018}.

There exists a close link between supergranulations and the fields that make up the quiet-region magnetic network and the active-region plages. Simultaneous multi-wavelength analyses of solar magnetographs in different spectral lines have shown that the quiet-region network and plages consist of the same type of small-scale magnetic structures (i.e., magnetic elements). A model of supergranulation network in the quiet-Sun and plages was derived by \cite{Stenflo1975} using the continuum, line profile, and EUV magnetograph data, which proposed a mechanism for the amplification of the magnetic field involving vortical motions around the downdrafts (sinks). It is well-known
that the magnetic-field concentrations are co-spatial with downdrafts in the quiet-Sun network \citep{Tanenbaum1969} and plages \citep{Howard1971}, similar to the whirl in a bathtub \citep{Norlund1985} and in an atmospheric cyclone. Due to the high electrical conductivity, the magnetic field is frozen into the plasma so the velocity vortex tube will wind up the magnetic field lines and generate electrical currents that may pinch the plasma column, resulting in the formation of high-field magnetic flux ropes. According to this model, efficient heating by hydromagnetic waves builds up an excess gas pressure
inside the twisted magnetic flux ropes located in the intergranular lanes which is released by the ejection of spicules into the corona. When an increasing magnetic flux is injected into a given area, the interaction between magnetic flux ropes and granulations yields changes in the structure of magnetic flux ropes and transition to pores and sunspots. Recent Hinode observation of photospheric turbulence in a plage and MHD simulation of magnetoconvective turbulence \citep{Chian2014} as well as the Hinode observation of supergranulation downflow vertices in the quiet-Sun \citep{Requerey2018} have confirmed that the downdraft is co-spatial with magnetic flux concentrations located at supergranular junctions, as pointed out by \cite{Tanenbaum1969}, \cite{Howard1971} and, \cite{Stenflo1975}. 

It was suggested by \cite{Stenflo1975} that persistent vortices tend to form near locations where the laminar flow is interrupted by obstacles in fluids and plasmas, e.g., in the Earth's atmosphere persistent vortices such as tropical cyclones are formed at locations where different pressure fronts collide. In supergranular junctions, the sinks in the NE are enhanced by the localized downdrafts where intense inhomogeneous magnetic fields and turbulent shearing motions produce numerous pressure fronts that interact with each other. Hence, the preferential sites in the solar photosphere for the formation of persistent velocity vortices and magnetic flux tubes are supergranular junctions where the flows from various neighboring supergranular cells meet and interact. The ideas of \cite{Stenflo1975} on the transport and evolution of magnetic fields and the formation of persistent vortices in the quiet Sun have been investigated and confirmed in detail by a number of papers based on high-resolution Hinode data \citep{Orozco2012, Giannattasio2013, Giannattasio2014a, Giannattasio2014b, Giannattasio2018, Caroli2015, Agrawal2018, Requerey2018}. \cite{Orozco2012} showed that magnetic elements are ubiquitous across the surface of the quiet Sun. They located the centre of the supergranule cell at the position of minimum radial velocity calculated from the magnetic element tracking in the time-averaged flow map. The computed horizontal velocity field of the IN magnetic elements is mostly radial and directed from the centre of a supergranulation to its boundaries with a non-constant velocity. On short timescales, the chaotic motions of magnetic elements are driven mainly by granular flows, but on long timescales they are driven by supergranular flows. In the NE, the combined action of inflows from neighboring supergranules traps magnetic elements in sinks. \cite{Giannattasio2013} studied the long-duration transport by tracking the turbulent convective flows of a large number of magnetic elements over 25 hr in the field of view of an entire supergranule cell and identified a double-regime behaviour in the displacement spectrum of the magnetic elements: superdiffusive in the whole field of view up to granular spatial scales ($\sim$ 1.5 Mm) and slightly diffusive up to supergranular spatial scales ($\sim$30$\sim$35 Mm).  \cite{Giannattasio2014a} investigated the separation spectrum for different values of the initial pair separation of magnetic elements for an entire supergranule and found out that the rate of pair separation depends on the spatial scale under consideration.  \cite{Giannattasio2014b} found that the amplitude of horizontal velocity field is small in the centre of the supergranular cell ($\sim$10 m/s), and increases toward the supergranular boundaries reaching a maximum ($\sim$600 m/s) at about half the radius. Under the assumption of magnetic elements passively transported by the flow, they confirmed double transport regimes in supergranules: superdiffusive in the IN, and a lower diffusivity regime in the NE which allows the amplification of magnetic fields in the supergranular junctions. In particular, they concluded that the IN  magnetic elements emerging from the centre of a supergranular cell have a maximum lifetime of $\sim$ 3 hr, not long enough to reach the supergranular boundaries to form the network. This finding indicates the need to find a physical mechanism that allows the transport of IN magnetic elements to the NE.    \cite{Gosic2014} used an automatic feature tracking algorithm to follow the evolution of IN and NE magnetic elements. They showed that IN magnetic elements continually transfer their flux to the NE by interacting with the NE magnetic elements via the merging (cancellation) processes that add (remove) flux. \cite{Caroli2015} investigated the advection-diffusion in photospheric flows by examining the evolution of the two-dimensional probability distribution function (PDF) of the mean-square displacements of magnetic elements. For times shorter than $\sim$ 2000 s the PDF seems to broaden symmetrically with time, but for longer times a multi-peaked PDF emerges which suggests the non-trivial nature of the transport process. They performed a Voronoi analysis to study the dynamics of convective structures using an automated polygon subdivision based on localization of magnetic elements. The resulting Voronoi diagram shows that the spatial distribution of magnetic elements is organized into several clumps embedded in regions of magnetic voids, related to the NE and IN regions. \cite{Gosic2016} extended the investigations of \cite{Iida2010} and \cite{Iida2012} to find out how small-scale IN magnetic fields appear and disappear on the solar surface. They showed that flux disappears from the IN through fading of magnetic elements, cancellation between opposite-polarity features, and interactions with the NE patches, which converts IN elements into NE features. \cite{Agrawal2018} used high cadence MURaM simulations to show that the observed superdiffusive scaling of the motions of IN magnetic elements at short increments is a consequence of random changes in barycentre positions due to flux evolution, and suggest that measurements of magnetic element motions must be used with caution in turbulent diffusion or wave excitation models. They proposed that instead of tracking magnetic elements over long periods of time, it may be preferable to compute the photospheric horizontal velocity using methods such as LCT to obtain more robust transport statistics. \cite{Giannattasio2018} applied the concepts of complex systems to compute the occurrence and persistence parameters in order to determine the characteristic spatial and temporal scales in which turbulent convection organises the magnetic features in the quiet Sun. The occurrence parameter is a measure of the tendency of a specific site to host magnetic features, whereas the persistence parameter quantifies the characteristic time range during which the same magnetic features linger in a specific location. They found that the occurrence parameter varies from $\sim$0\% in the central part of the supergranular cell to $\sim$100\% near its vertices. Their persistence analysis identified long-correlated patterns in the supergranular boundaries where decorrelation times up to $\sim$1240 min are found, indicating that on average the NE magnetic elements move slower and are correlated longer in relation to the IN magnetic elements. \cite{Requerey2018} studied the relation between a persistent photospheric vortex flow and the evolution of a NE magnetic element at a supergranular vertex. Supergranular cells are identified as large-scale divergence structures in the flow map. At their vertices and co-spatial with an NE magnetic element, the horizontal velocity flows converge to a central point. One of these converging flows is identified as a persistent vortex during the entire 24 h time series. It consists of three consecutive vortices with a lifetime of $\sim$7 h each that are seen almost at the same location. At their core, an NE magnetic element is also seen. Its evolution is strongly correlated to that of the vortices, namely, the magnetic feature is concentrated and evacuated when it is caught by the vortices and is weakened and fragmented after the whirls disappear. Such behavior is consistent with the theoretical picture of the stabilisation of a magnetic flux tube by a surrounding vortical flow \citep{Schussler1984}.

The quiet Sun \citep{Bellot2019} provides a unique laboratory for studying the fundamental characteristics of astrophysical magnetism and turbulence thanks to recent advances in space and ground observations, numerical simulations, and nonlinear dynamics theories. In particular, the newly developed tool of Lagrangian coherent structures (LCS) \citep{Haller2000,Shadden2005,Mathur2007,Haller2015} provides an efficient method to uncover the Lagrangian skeleton of turbulence. Due to the Lagrangian nature, LCS behave as a separatrix of particles in the fluid by forming contours to separate regions of different particle movements. Hence, LCS are distinguished material lines/surfaces that predominantly influence the neighboring particle trajectories in a given time interval by organising the flow into coherent patterns. There are various definitions of LCS depending on whether there is greater repulsion/attraction (hyperbolic LCS), shear (parabolic LCS), or swirling motions (elliptic LCS) of particles in the fluid for the considered time interval. In this paper, we will focus on the investigation of hyperbolic repelling/attracting LCS and elliptic LCS in solar photosphere.

We will demonstrate that the study of LCS in supergranular turbulence renders a strong support for the model of supergranulation network of \cite{Stenflo1975} and the ground and space-borne observations of supergranulations in a quiet-Sun. The nonlinear dynamics tools of LCS have been applied to study turbulence in terrestrial atmosphere \citep{Rutherford2015}, jovian atmosphere \citep{Hadjighasem2017}, ocean \citep{Beron-Vera2018}, nuclear fusion \citep{Padberg2007, Borgogno2011} and astrophysical plasmas \citep{Rempel2011, Rempel2012, Rempel2013, Rempel2017, Yeates2012, Chian2014, Silva_2018b}. A series of papers applied the technique of finite-time Lyapunov exponent to detect hyperbolic LCS in numerical simulation of MHD turbulence in ABC dynamo \cite{Rempel2011, Rempel2012, Rempel2017}, $\alpha^2$ mean-field dynamo \citep{Rempel2011, Rempel2013}, and large-scale dynamo in turbulent compressible convection with uniform horizontal shear and rotation \citep{Chian2014}. \cite{Yeates2012} used the horizontal velocity field derived from the line-of-sight magnetogram of Hinode/SOT to study LCS in a plage of solar active region AR 10930. They showed that the network of quasi-separatrix layers \citep{Demoulin1996} in the magnetic field corresponds to the repulsive LCS, and demonstrated how to infer the build-up of magnetic gradients in the solar corona directly from the photospheric horizontal velocity by calculating the magnetic connectivity measures such as the squashing factor without recourse to magnetic field extrapolation. \cite{Chian2014} used the same dataset of \cite{Yeates2012} to show that on average the deformation Eulerian coherent structures dominate over vortical Eulerian coherent structures in a plage. They used Hinode data and MHD simulation of magnetoconvection to demonstrate that the network of high magnetic flux concentration in the intergranular lanes corresponds to the attracting LCS. \cite{Rempel2017} extended the technique for detecting objective velocity vortices developed by \cite{Haller2016} to objective magnetic vortices. \cite{Silva_2018b} applied the LAVD technique of \cite{Haller2016} and the Hinode dataset of \cite{Requerey2018} to detect objective velocity vortices in the quiet-Sun, and introduced a d-criterion to avoid false vortex detection.

The aim of this paper is to unravel the Lagrangian skeleton of supergranular turbulence by detecting hyperbolic and elliptic Lagrangian coherent structures in a quiet Sun in Sec. 3,  based on high-resolution Hinode observation of solar photosphere described in Sec. 2. Discussions and conclusion are given in Sec. 4.

\section{PHOTOSPHERIC FLOWS AND MAGNETIC FIELDS IN A QUIET SUN}
 \subsection{ Photospheric flows}
Our data analysis is based on the magnetograms of the disk center of a quiet Sun captured by the Narrowband Filter Imager (NFI) onboard the Hinode satellite on 2010 November 2, within the framework of the Hinode Operation Plan 151 entitled ``Flux replacement in the photospheric network and internetwork", with a cadence of 90 s, a field-of-view of 80" x 74", a pixel size of 0".16 and a spatial resolution of 0".3 \citep{Gosic2014}. 

\begin{figure*}
    \centering
     \subfigure[][\label{fig:1a}]{\includegraphics[trim =1mm 1mm 1mm 1mm, clip,width=0.48\textwidth]{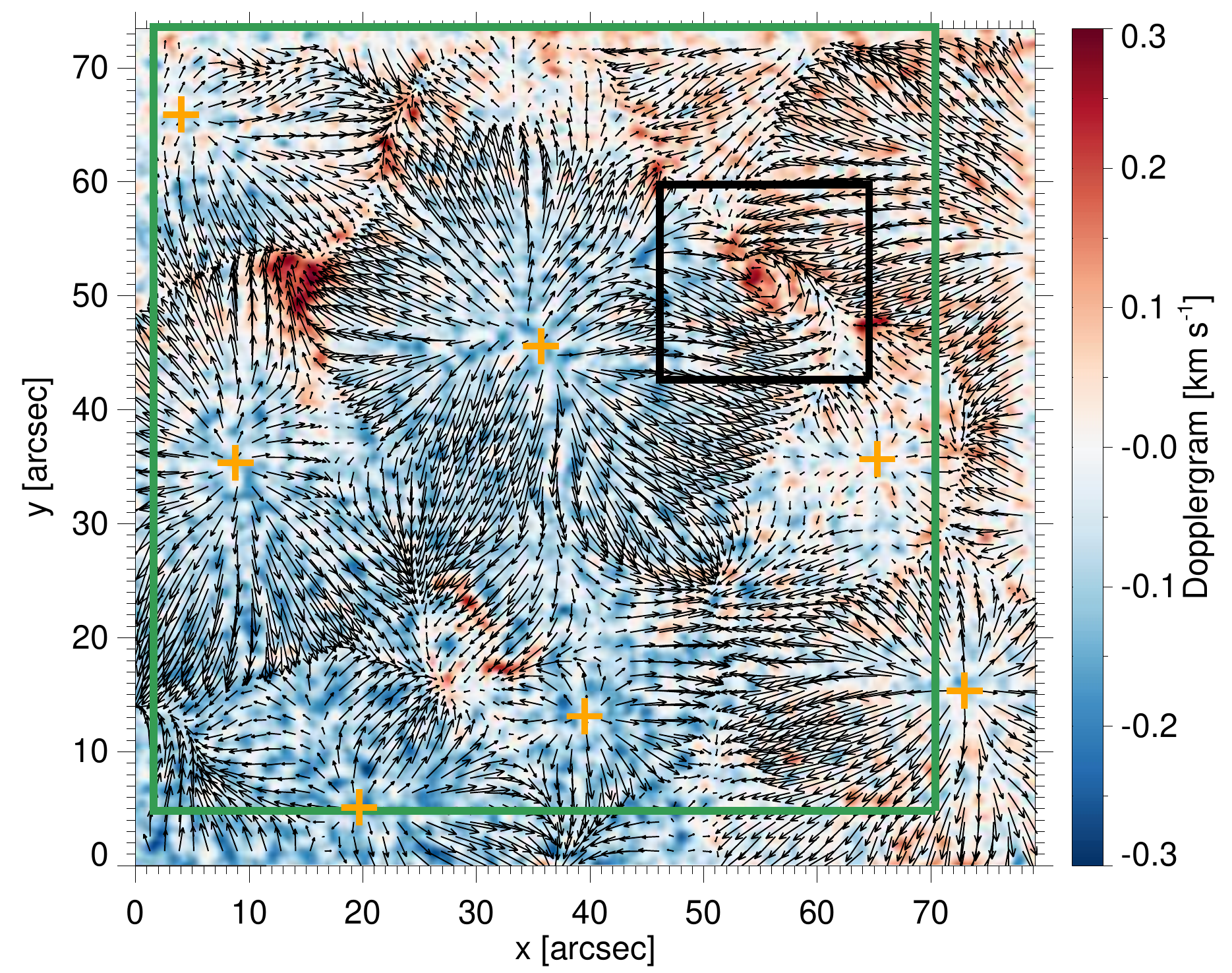}}
\qquad
    \subfigure[][\label{fig:1b}]{\includegraphics[trim =1mm 1mm 1mm 1mm, clip,width=0.48\textwidth]{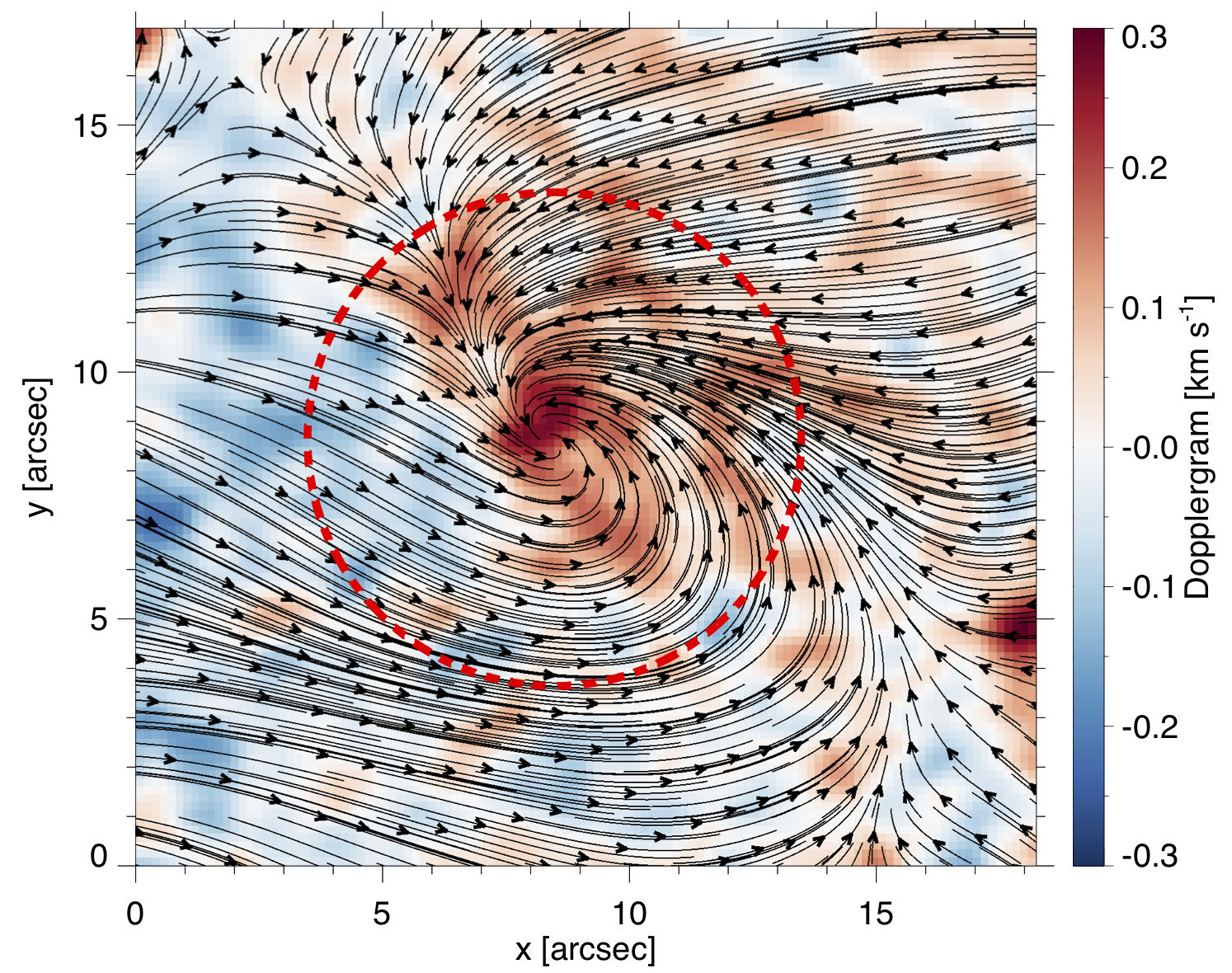}}
    \caption{Horizontal photospheric flow, Dopplergram, and persistent vortex observed by Hinode at the disk centre of a quiet-Sun on 2010 November 2. The photospheric horizontal velocity field as deduced by LCT (black arrows) and the mean Dopplergram in the background (where red and blue regions correspond to downflow and upflow areas, respectively), both time-averaged from 16:46:26 UT to 23:46:34 UT. Each orange cross denotes
the mean centre of a supergranular cell determined by the local minimum of the mean horizontal velocity field. The black rectangle denotes the location of a persistent vortex. The green rectangle indicates the area where the study of Lagrangian coherent structures will be carried out in this paper. (b) An enlarged view of the region marked by the black rectangle in (a), showing that the persistent vortex is located at a site where photospheric flows from the neighboring supergranular cells meet and interact.} 
    \label{fig:1}
\end{figure*}

Supergranular cells are detected using Dopplergram time sequences by dividing them in 2-hr intervals and applying the Local Correlation Tracking (LCT) technique to each sequence to determine the horizontal velocity field \citep{November1988,Gosic2014}. In Fig. \ref{fig:1}\subref{fig:1a} the black arrows display the mean horizontal velocity field for 7-hr interval from 16:46:26 UT
to 23:46:34 UT; the background image displays the time-averaged Dopplergram of the line-of-sight velocities for the same time interval. In the centre of the map, a large supergranular cell is seen with a divergence structure of $\sim$35 Mm x 35 Mm, surrounded by other supergranular cells. The horizontal velocities are directed radially outward from the mean centre (marked by an orange cross where lies the local minimum of the mean horizontal velocity field) of each supergranular cell toward its boundary. In the interior of a supergranular cell, the Dopplergram is dominated by upflows, while the supergranular boundary is dominated by downflows. A particularly prominent localized downflow in a supergranular vertex is marked by a black rectangle. 
A zoom in the flow streamlines of this area given in Fig. \ref{fig:1}\subref{fig:1b} shows spiraling velocity streamlines. Therefore, the dynamics of this region is affected by vortical motions during the time interval used to compute the mean velocity field. The  centre of this persistent vortex with a mean diameter of $\sim$5 Mm is located in the centre of downdraft, which is indicative of a flow driven by the bathtub effect. \cite{Requerey2018} studied this vortical structure and showed that an NE magnetic element is detected by the magnetogram at the centre of the vortex flow, which co-rotates with the vortex in the counter-clockwise direction. The vortex seen in the magnetogram has a diameter of $\sim$10 Mm which is larger than the bright point counterpart seen in the intensity map \citep{Requerey2018}. Note that flows from other nearby supergranular cells converge and interact in this area. Similar photospheric vortex flows at supergranular junctions have been detected previously by \cite{Brandt1988} and \cite{Attie2009}.

\begin{figure*}
\center
\subfigure[ref1][\label{fig:2a}]{\includegraphics[trim =1mm 10mm 1mm 10mm, clip,width=0.48\textwidth]{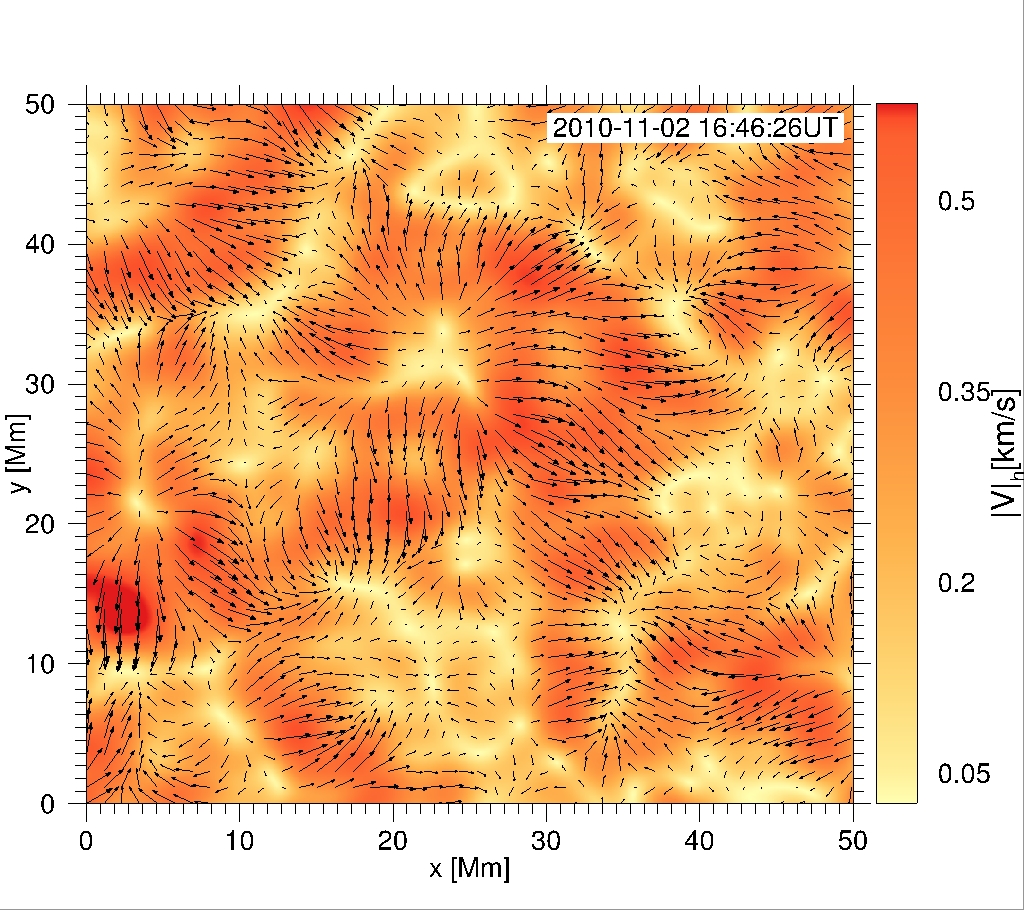}}
\qquad
\subfigure[ref1][\label{fig:2b}]{\includegraphics[trim =1mm 10mm 1mm 10mm, clip,width=0.48\textwidth]{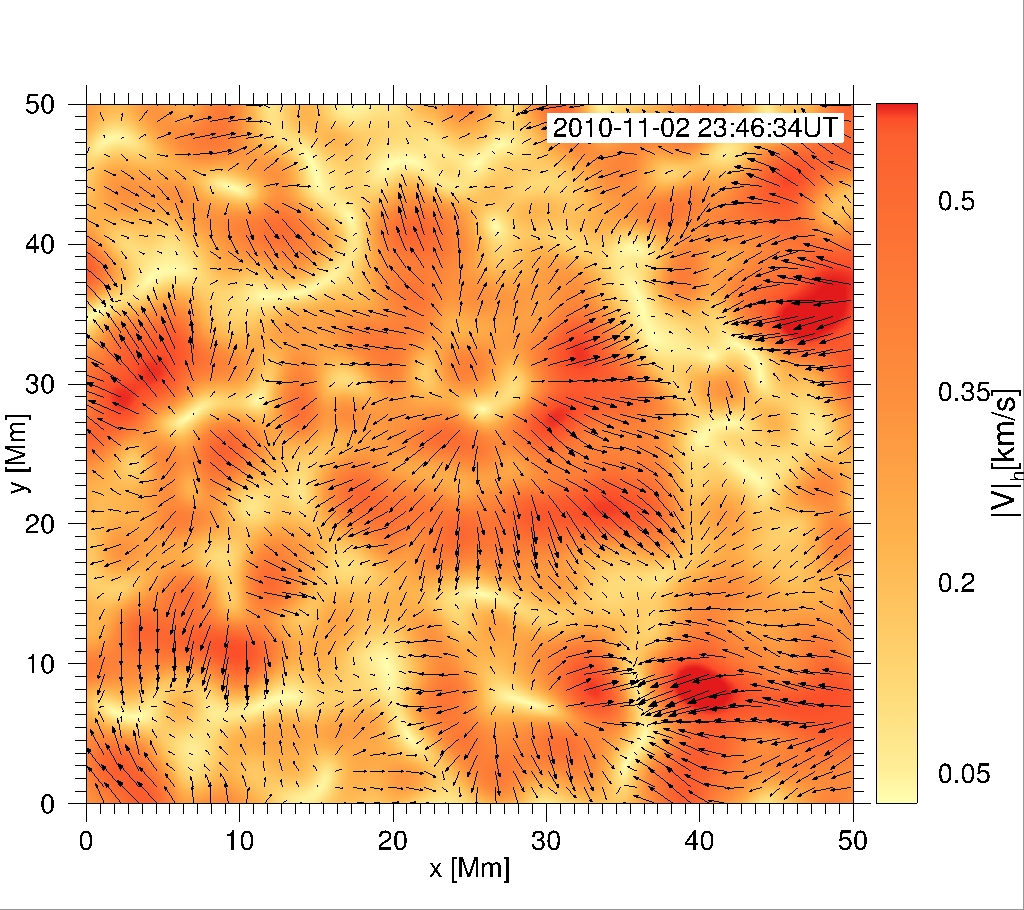}}
\qquad
\subfigure[ref1][\label{fig:2c}]{\includegraphics[trim =1mm 5mm 1mm 5mm, clip,width=0.48\textwidth]{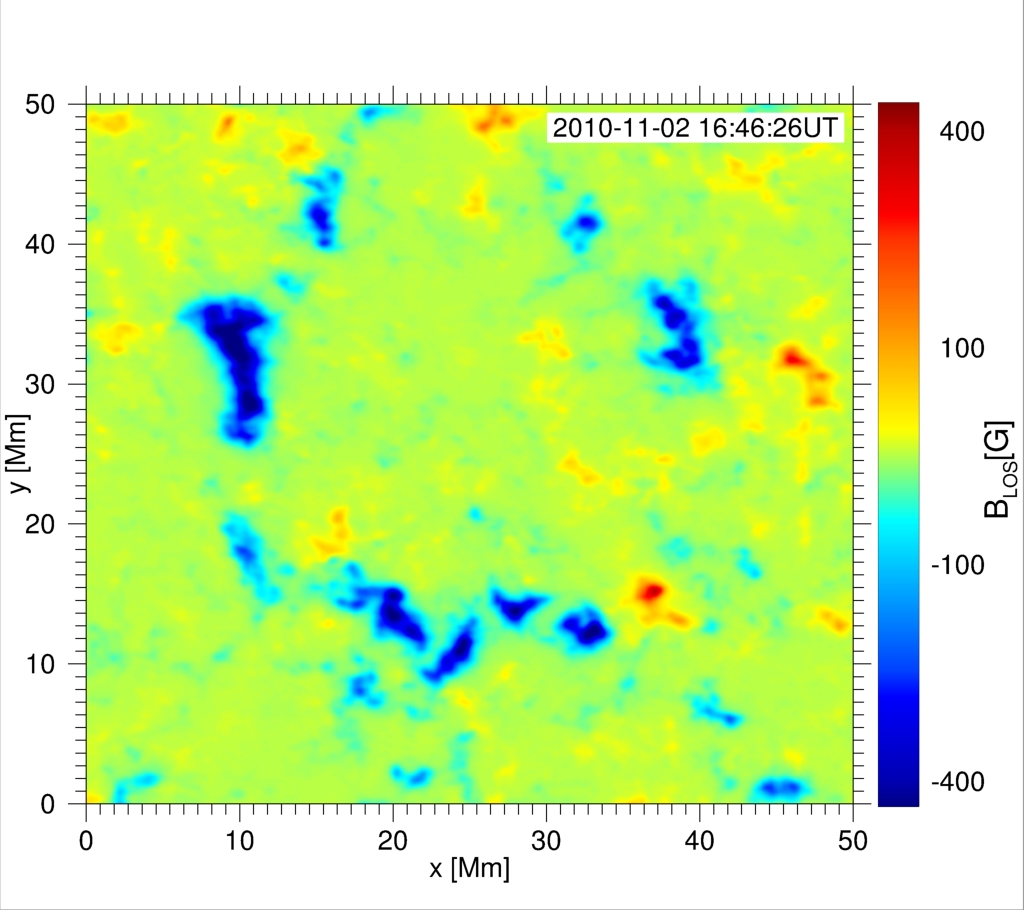}}
\qquad
\subfigure[ref1][\label{fig:2d}]{\includegraphics[trim =1mm 5mm 1mm 5mm, clip,width=0.48\textwidth]{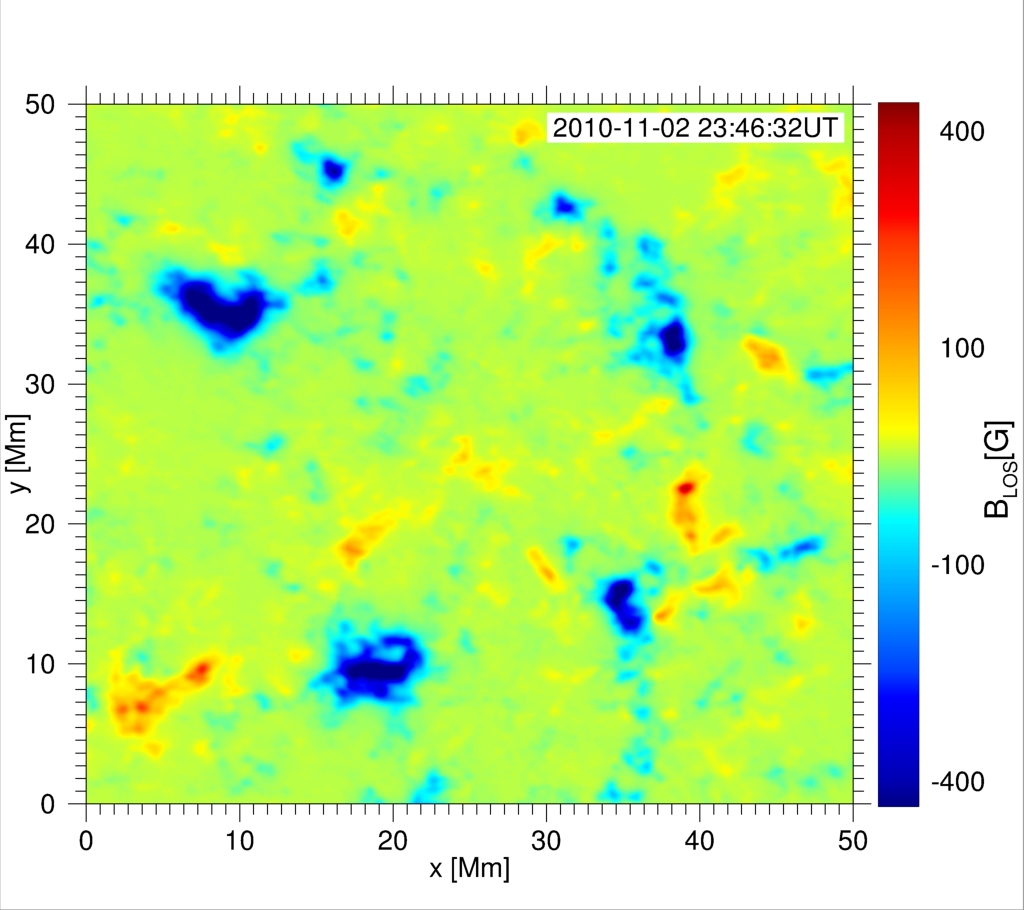}}
    \caption{Horizontal velocity field and line-of-sight magnetic field. The 2-hourly time-averaged horizontal velocity field (black arrows) deduced by applying LCT to the intensity maps and the horizontal velocity modulus (the background image) for the green rectangle region marked in Fig. \ref{fig:1}\subref{fig:1a} at 16:46:26 UT (a) and 23:46:34 UT (b). The corresponding 2-hourly time-averaged line-of-sight magnetic field at
16:46:26 UT (c) and 23:46:34 UT (d).}
    \label{fig:2}
\end{figure*}

In this paper, we detect hyperbolic and elliptic Lagrangian coherent structures using a time sequence of 281 frames of 2-hourly time-averaged horizontal velocity field deduced by LCT from the intensity maps for the green box region marked in Fig. \ref{fig:1}\subref{fig:1a} during an interval of 7 hours from 16:46:26 UT to 23:46:34 UT. Figures \ref{fig:2}\subref{fig:2a} and \ref{fig:2}\subref{fig:2b} show the 2D horizontal velocity field at 16:46:26 UT and 23:46:34 UT, respectively. In Figs. \ref{fig:2}\subref{fig:2c} and \ref{fig:2}\subref{fig:2d}, we plot the corresponding 2-hourly time-averaged images of the line-of-sight magnetic field. The boundary of supergranular cells are given by the regions of low values of the mean horizontal velocity modulus, corresponding to the regions of high values of the mean downflow velocity in the Dopplergram of Fig. \ref{fig:1}\subref{fig:1a}. It follows from Fig. \ref{fig:2} that there are considerable changes in both velocity- and magnetic-field distributions during the time interval of 7 hours.

\subsection{Photospheric magnetic fields}
\begin{figure*}
    \centering
     \subfigure[][\label{fig:3a}]{\includegraphics[trim =1mm 50mm 1mm 50mm, clip,width=0.48\textwidth]{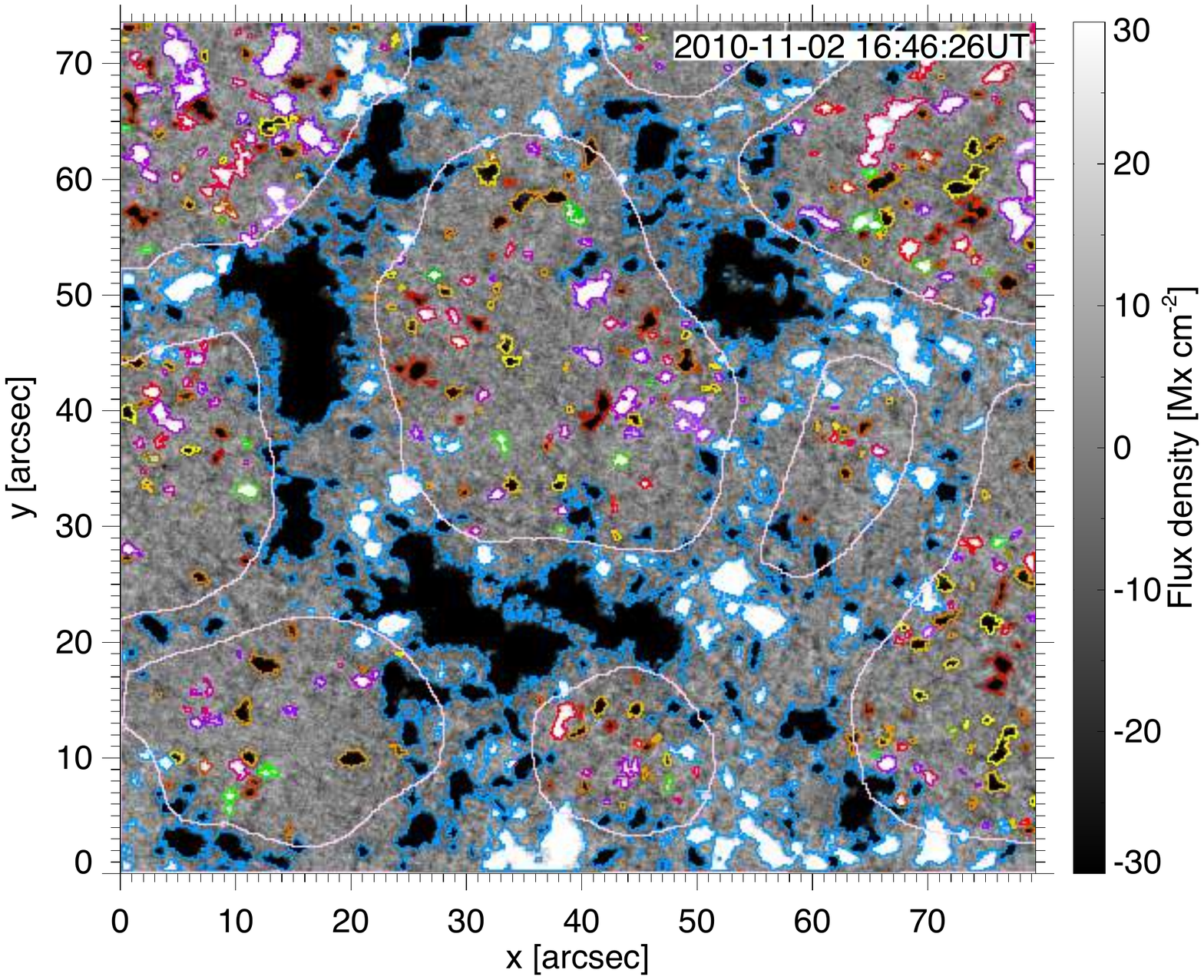}}
\qquad
    \subfigure[][\label{fig:3b}]{\includegraphics[trim =1mm 50mm 1mm 50mm, clip,width=0.48\textwidth]{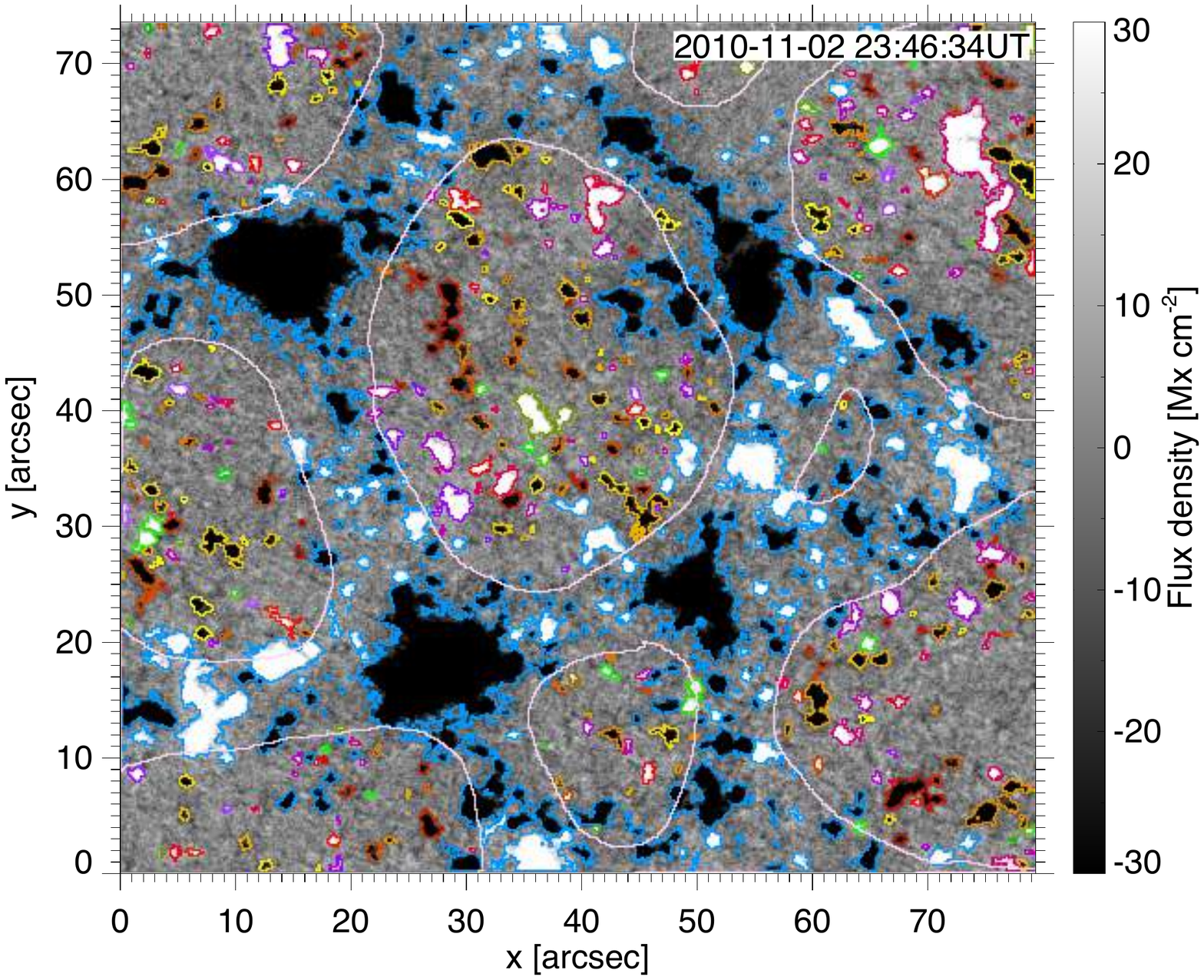}}
    \caption{Snapshot magnetograms of supergranular cells. Snapshot of the original magnetogram at 16:46:26 UT (a) and 23:46:34 UT (b). The pink contours outline the Eulerian (snapshot) boundaries of supergranular cells that separate NE and IN regions as deduced by LCT; blue contours indicate the NE magnetic elements; IN magnetic elements are displayed in other colors. The background images display snapshot of the magnetic flux density at 16:46:26 UT (a) and 23:46:34 UT (b), respectively.}
    \label{fig:3}
\end{figure*}

We apply the YAFTA code \citep{Welsch2003} to detect and track magnetic patches in magnetograms by assigning a unique label to each feature of magnetic elements which are in constant motion advected by the photospheric flow and often interact with other features, gaining or losing flux, until they eventually disappear from the solar surface. The following dynamical processes need to be considered to track their evolution: in situ appearance/disappearance, merging, fragmentation and cancellation \citep{Gosic2014}. Figures \ref{fig:3}\subref{fig:3a} and \ref{fig:3}\subref{fig:3b} display snapshots of magnetograms at 16:46:26 UT and 23:46:34 UT, respectively, for the photospheric region of Fig. \ref{fig:1}\subref{fig:1a}. Pink contours outline the boundaries of supergranular cells as deduced from LCT, which separate NE and IN regions. Blue contours indicate the NE elements, and IN elements are indicated in other colors. These magnetograms show IN flux patches, many of which eventually supply flux to the NE. 

The background images in Figs. \ref{fig:3}\subref{fig:3a} and \ref{fig:3}\subref{fig:3b} show an Eulerian (snapshot) view of the magnetic flux density at 16:46:26 UT (a) and 23:46:34 UT (b), respectively. It follows from Figs. \ref{fig:1} and \ref{fig:3} that as the IN magnetic elements are advected by the radially directed outflows, from the centre to the boundary of a supergranular cell, they undergo a rich variety of dynamical processes. In particular, when IN magnetic elements reach the supergranular boundary they interact with the NE magnetic elements which allows the deposit of their flux in supergranular junctions \cite[for more details see][]{Gosic2014, Gosic2016}. Furthermore, the shape and size of supergranular cells as well as the morphology and structure of the NE in supergranular junctions vary considerably in 7 hours from Fig. \ref{fig:3}\subref{fig:3a} to Fig. \ref{fig:3}\subref{fig:3b}.

\section{LAGRANGIAN COHERENT STRUCTURES IN SUPERGRANULAR TURBULENCE}
\subsection{Hyperbolic Lagrangian coherent structures}

First, we determine the hyperbolic Lagrangian coherent structures in supergranular turbulence by computing the forward/backward finite-time Lyapunov exponent (f-FTLE/b-FTLE) \citep{Shadden2005,Padberg2007,Borgogno2011,Rempel2011, Rempel2012, Yeates2012,Rempel2013, Chian2014} of the horizontal velocity field by advecting a dense grid of $405 \times 405$ tracer particles over the domain of interest. Haller et al. (2015) pointed out that the FTLE is not objective, i.e., invariant under time-dependent translations and rotations of the reference frame. Hence, the FTLE results must be handled with caution. Consider a tracer particle advected by the velocity field $\boldsymbol{u}(\boldsymbol{r}, t)$ from an initial time $t_0$ and solve the particle advection equation 
\begin{equation}
\label{Eq:1}
\centering
    \frac{d \boldsymbol{r}}{dt}= \boldsymbol{u}(\boldsymbol{r},t) 
\end{equation}
over a grid of initial positions $\boldsymbol{r}_0$ until the final positions $\boldsymbol{r}(t_0 + \tau)$ are reached after a finite-time duration $\tau$. The FTLEs of the particle trajectories for a 2D flow are calculated at each initial position $\boldsymbol{r}_0$ as
\begin{equation}
\label{Eq:2}
    \sigma_i ^{t_0+\tau} (\boldsymbol{r}_0)= \frac{1}{|\tau|}\ln{\sqrt{\lambda_i}}, \hspace{5mm} i=1,2
\end{equation}
where $\lambda_i$ ($\lambda_1 >\lambda_2$) are the eigenvalues of the finite-time right Cauchy-Green deformation tensor $\Delta = J^{\top}J$, in which $J= d\phi^{t_0+\tau}_{t_0}(\textbf{r})/d\textbf{r}$ is the deformation gradient, $\top$ denotes the transpose, and $\phi^{t_0+\tau}_{t_0}: \boldsymbol{r}(t_0)\rightarrow \boldsymbol{r}(t_0+\tau)$ is the flow map for Eq. (\ref{Eq:1}). Advecting a particle forward in time reveals the repelling LCS in the f-FTLE field, which is the source of stretching in the flow, whereas advecting a particle backward in time reveals the attracting LCS in the b-FTLE field along which particles congregate to form the observable patterns. Equation (\ref{Eq:1}) is solved using a fourth-order Runge-Kutta integrator; to obtain a continuous time-varying velocity field, cubic splines are used for space and time interpolations using a discrete set of velocity field frames previously obtained with the LCT method.

\begin{figure*}
\center
\subfigure[ref1][\label{fig:4a}]{\includegraphics[trim =1mm 10mm 1mm 10mm, clip,width=0.48\textwidth]{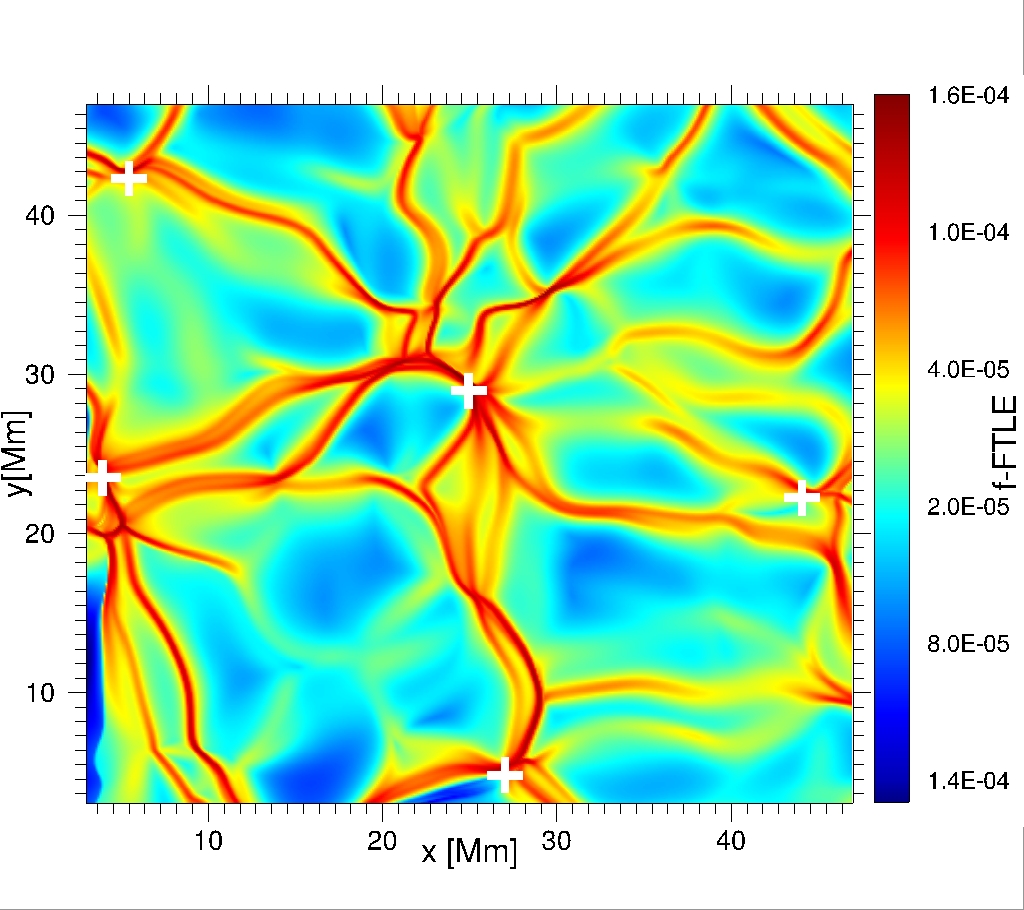}}
\qquad
\subfigure[ref1][\label{fig:4b}]{\includegraphics[trim =1mm 10mm 1mm 10mm, clip,width=0.48\textwidth]{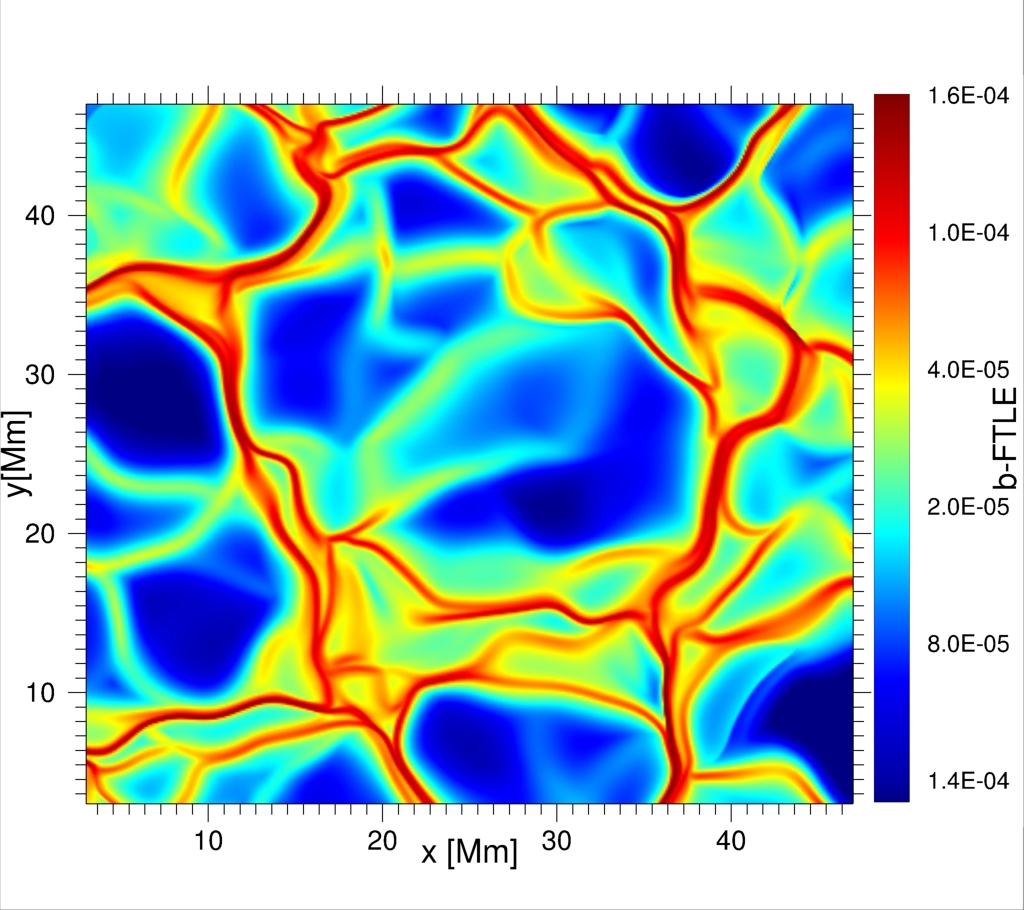}}
\qquad
\subfigure[ref1][\label{fig:4c}]{\includegraphics[trim =1mm 5mm 1mm 5mm, clip,width=0.48\textwidth]{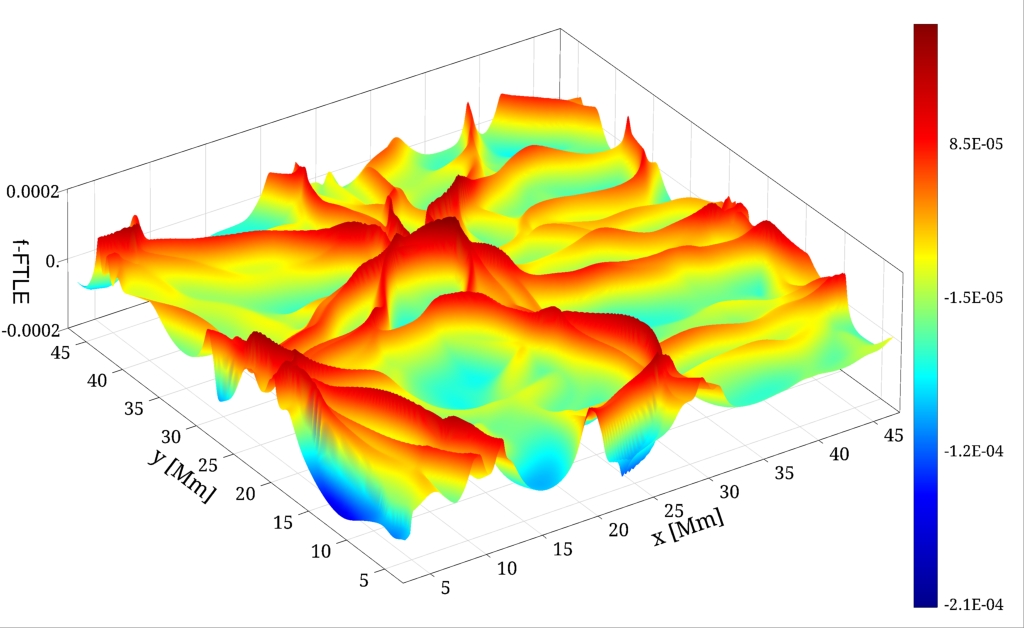}}
\qquad
\subfigure[ref1][\label{fig:4d}]{\includegraphics[trim =1mm 5mm 1mm 5mm, clip,width=0.48\textwidth]{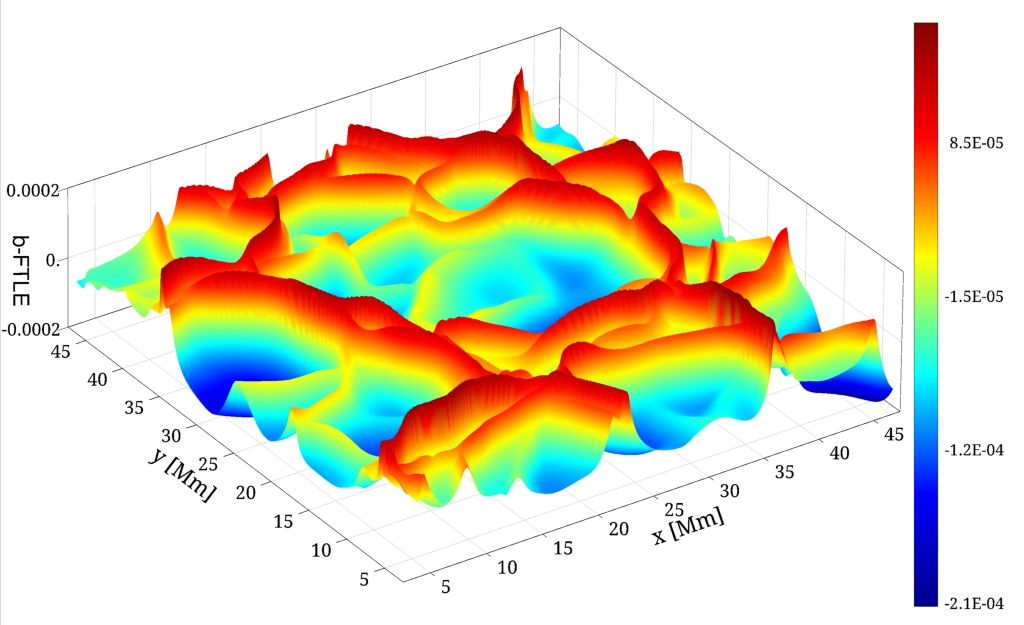}}
    \caption{Hyperbolic repelling and attracting Lagrangian coherent structures of supergranular turbulence in the time interval from 16:46:26 UT to 23:46:34 UT on 2010 November 2. (a) 2D plot of the f-FTLE computed for $t_0$= 16:46:26 UT and $\tau$ = +7h. (b) 2D plot of the b-FTLE computed for $t_0$= 23:46:34 UT and $\tau$ = -7h. (c) 3D surface-plot of the f-FTLE of (a). (d) 3D surface-plot of the b-FTLE of (b). The white crosses in (a) indicate the Lagrangian centres of supergranular cells determined by the local maxima of the f-FTLE. Lagrangian centres of nearby supergranular cells are interconnected by ridges of the f-FTLE. Lagrangian boundaries of supergranular cells are given by ridges of the b-FTLE.}
    \label{fig:4}
\end{figure*}

Figure \ref{fig:4}\subref{fig:4a} shows a 2D plot of f-FTLE for the supergranular region of Fig. \ref{fig:1}\subref{fig:1a}  marked by the green rectangle, computed forward in time using the sequence of 281 frames of the horizontal velocity field from 16:46:26 UT to 23:46:34 UT. The white crosses mark the locations of the time-dependent Lagrangian centres of  supergranular cells characterised by local maxima of f-FTLE. Thin ridges of large positive f-FTLE in Fig. \ref{fig:4}\subref{fig:4a}  represent the locally strongest repelling material lines which exert the most influential impact on the diverging transport of supergranular flows in the given time interval. \cite{Yeates2012} and \cite{Chian2014} demonstrated that f-FTLE is closely related to the squashing Q-factor, which identifies the most likely locations for the occurrence of magnetic reconnection \citep{Demoulin1996, Inoue2016}. Hence, the ridges of f-FTLE indicate the preferential sites for magnetic elements to interact and reconnect. 

Figure \ref{fig:4}\subref{fig:4b} shows a 2D plot of b-FTLE for the supergranular region of Fig. \ref{fig:1}\subref{fig:1a}  marked by the green rectangle, computed backward in time using the sequence of 281 frames of the horizontal velocity field from 23:46:34 UT to 16:46:26 UT. In Figs. \ref{fig:3}\subref{fig:3a} and  \ref{fig:3}\subref{fig:3b} we get a rough (estimated) view of the Eulerian (snapshot) boundaries of supergranular cells at 16:46:26 UT and 23:46:34 UT, respectively. In contrast, Fig. \ref{fig:4}\subref{fig:4b}  gives an accurate view of the complex time-dependent Lagrangian boundaries of supergranular cells. Thin ridges of large positive b-FTLE in Fig. \ref{fig:4}\subref{fig:4b} represent the locally strongest attracting material lines that exert the most influence on the converging transport of supergranular flows. Note that the b-FTLE is closely related to the Lagrangian analysis of corks \citep{November1988, Roudier2018}, but it enables us to acquire an additional important information on the local rate of the flow convergence indicated by the ridges of large (positive) b-FTLE. \cite{Chian2014} established the correspondence of the network of high magnetic flux concentration to the attracting LCS, as seen in supergranular junctions.  Figures \ref{fig:4}\subref{fig:4c} and \ref{fig:4}\subref{fig:4d} exhibit surface plots of f-FTLE and b-FTLE, respectively, which present a 3D perspective of the thin ridges of the repelling and attracting LCS. The ridges of the repelling LCS seen in \ref{fig:4}\subref{fig:4c} provide the transport barriers that facilitate the concentration of strong magnetic fields and formation of vortices in the valleys (i.e., regions of low-value f-FTLE) of the repelling LCS. The ridges of the attracting LCS seen in \ref{fig:4}\subref{fig:4d} act as the sinks for downdraft of photospheric flows that lead to vortical motions. 

\begin{figure*}
    \centering
     \subfigure[][\label{fig:5a}]{\includegraphics[trim =1mm 1mm 1mm 1mm, clip,width=0.48\textwidth]{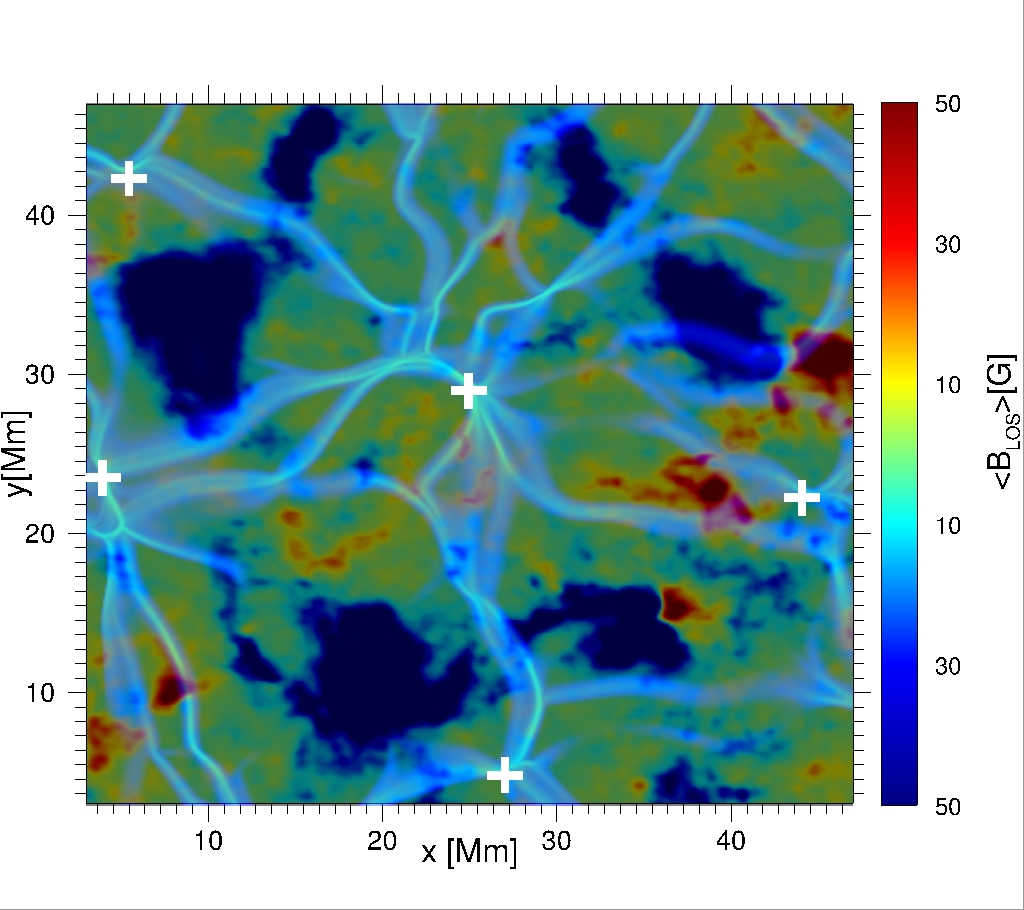}}
\qquad
    \subfigure[][\label{fig:5b}]{\includegraphics[trim =1mm 1mm 1mm 1mm, clip,width=0.48\textwidth]{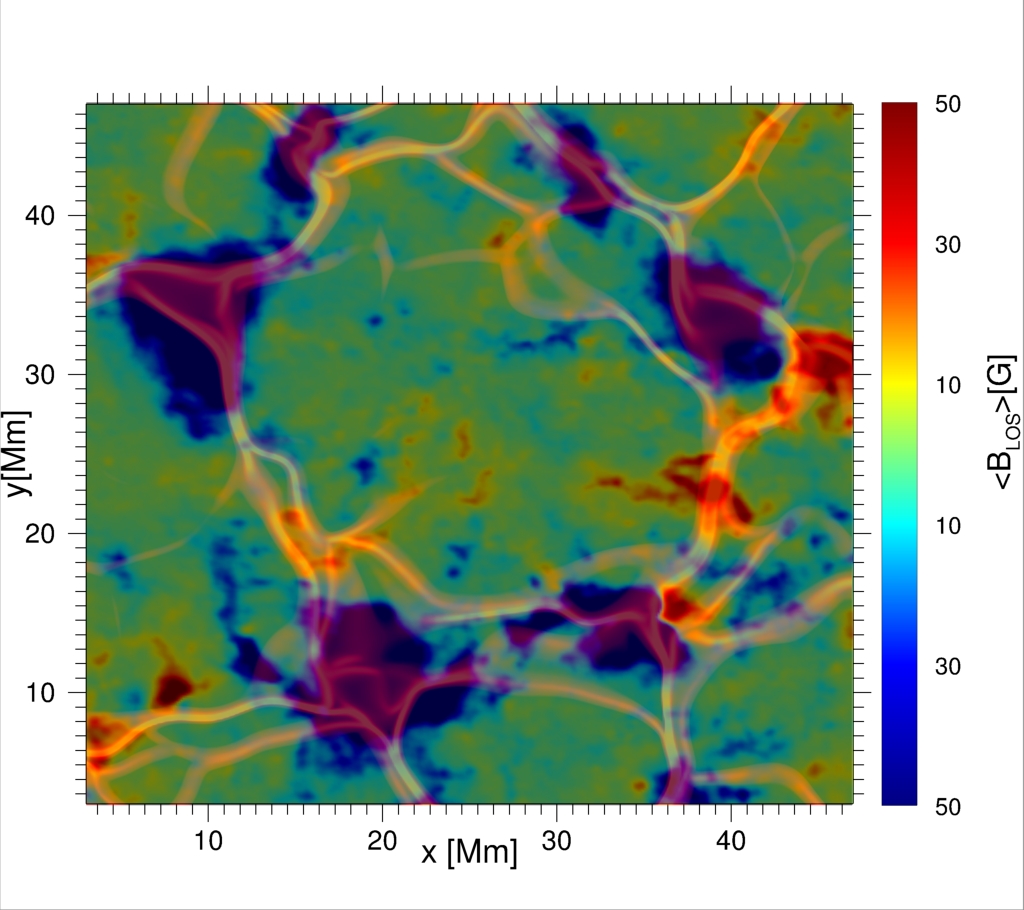}}
    \caption{Hyperbolic Lagrangian coherent structures and magnetogram. Superposition of hyperbolic repelling (blue material line, a) and attracting (orange material line, b) LCS with the background mean magnetogram of the line-of-sight magnetic field of supergranular cells time-averaged from t=16:46:26 UT to 23:46:34 UT. The white cross marks the Lagrangian centre of a supergranular cell.}
    \label{fig:5}
\end{figure*}

The role of hyperbolic LCS on the magnetic field distribution is revealed by Fig. \ref{fig:5} where we plot a superposition of the hyperbolic repelling (Fig. \ref{fig:5a}) and attracting (Fig. \ref{fig:5b}) LCS with the background time-averaged line-of-sight magnetic field from t=16:46:26 UT to 23:46:34 UT. Figure \ref{fig:5}\subref{fig:5a} shows clearly that the patches of intense mean magnetic fields in supergranular junctions are bounded by the ridges of high-value f-FTLE fields. Figure \ref{fig:5}\subref{fig:5b} shows clearly that the concentration of intense mean magnetic fields are aligned along the ridges of high-value b-FTLE fields. 

\subsection{Elliptic Lagrangian coherent structures}
The Lagrangian-averaged vorticity deviation (LAVD) is an objective method for detecting elliptic LCS (vortices) developed by \cite{Haller2016} for fluids. \cite{Rempel2017} and \cite{Silva_2018b} applied LAVD to detect vortices in plasmas. The LAVD field is computed by the integral
\begin{equation}
\label{Eq:3}
\mbox{LAVD}_{t_0}^{t0+\tau} := \int_{t_0}^{t_0+\tau}|\boldsymbol{\omega}(\boldsymbol{r}(s),s) -\langle \boldsymbol{\omega}(s) \rangle| ds
\end{equation}
where $\boldsymbol{\omega} = \nabla \times \boldsymbol{u}$ is the vorticity, $\tau$ denotes a finite time interval, and $\langle \cdot \rangle$ denotes the instantaneous spatial average. Fluid particle trajectories are determined by solving Eq. (\ref{Eq:1}) over a grid of initial positions $\boldsymbol{r}_0$ until the final positions $\boldsymbol{r}(t_0 + \tau)$ are reached after a finite time duration $\tau$.

For a finite time duration [$t_0$, $t_0$ + $\tau$], a Lagrangian vortex is defined as an evolving material domain filled with a nested family of convex tubular level surfaces of LAVD$_{t_0}^{t_0+\tau}$ with outward-decreasing LAVD values. In contrast to the Eulerian methods that detect instantaneous swirl-like structures that do not convey long-term information of fluid motions, it was demonstrated that the LAVD method is capable of detecting persistent vortical flows \citep{Haller2016,Rempel2017,Silva_2018b}. For 2D incompressible, shearless, non-magnetized fluids, the center of objective vortices are usually co-spatial with a local maximum of the LAVD field \citep{Haller2016}. This is not necessarily true for highly compressible magnetized fluids with strong shear regions such as photospheric flows, for which we must allow some deviation from convexity to ensure that all vortex cores are captured.  Moreover, \cite{Silva_2018b} showed that in order to avoid false vortex detection in flows with strong shear, such as supergranulations, it is necessary to apply a post-processing filter to the set of vortices detected by LAVD. The d-criterion, based on the geometry of the streamlines of the displacement vector field of fluid elements, is adopted to
find vortex cores. A grid point is considered a vortex core if the displacement vectors computed from its neighboring points indicate a clockwise or counter-clockwise motion pattern. The displacement vectors are computed by advecting the neighboring particles for a given time and subtracting their initial positions from their final positions. For details, see \citet{Silva_2018b}. A LAVD vortex is considered a true vortex boundary if it surrounds a vortex core detected by the d-criterion. 

\begin{figure*}
\center
\subfigure[ref1][\label{fig:6a}]{\includegraphics[trim =1mm 10mm 1mm 10mm, clip,width=0.48\textwidth]{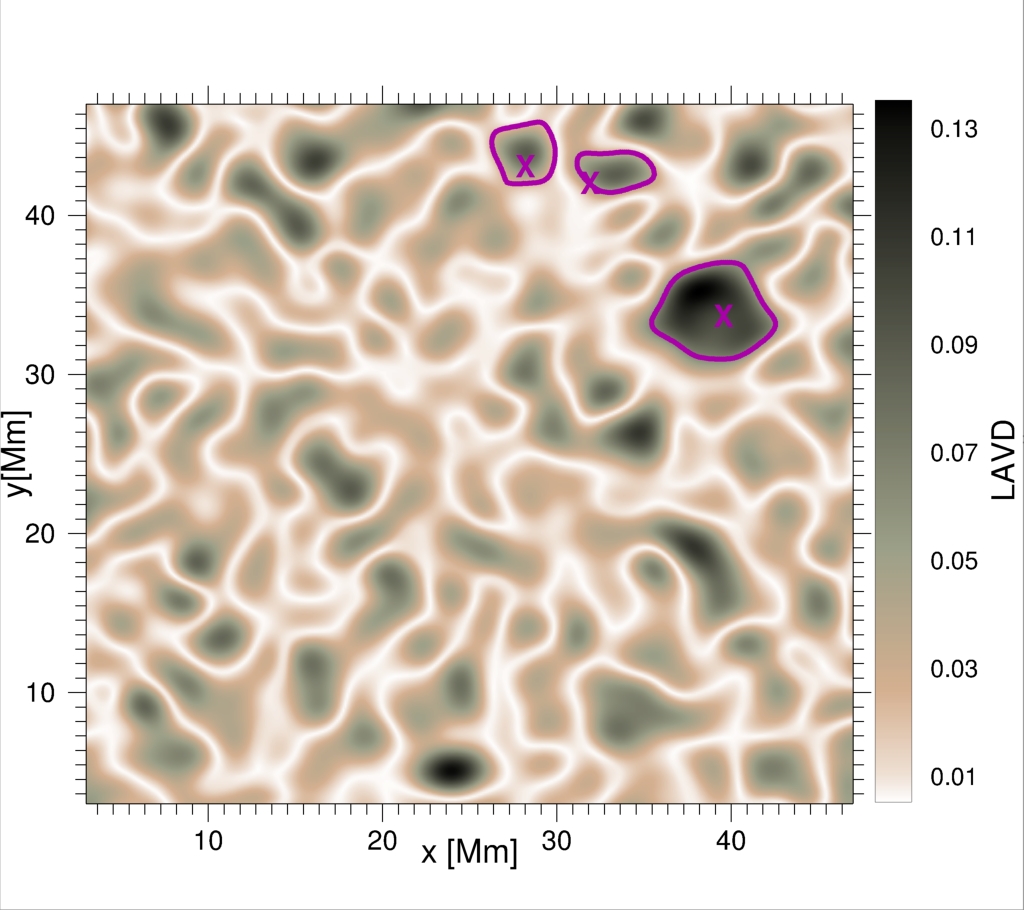}}
\qquad
\subfigure[ref1][\label{fig:6b}]{\includegraphics[trim =1mm 10mm 1mm 10mm, clip,width=0.48\textwidth]{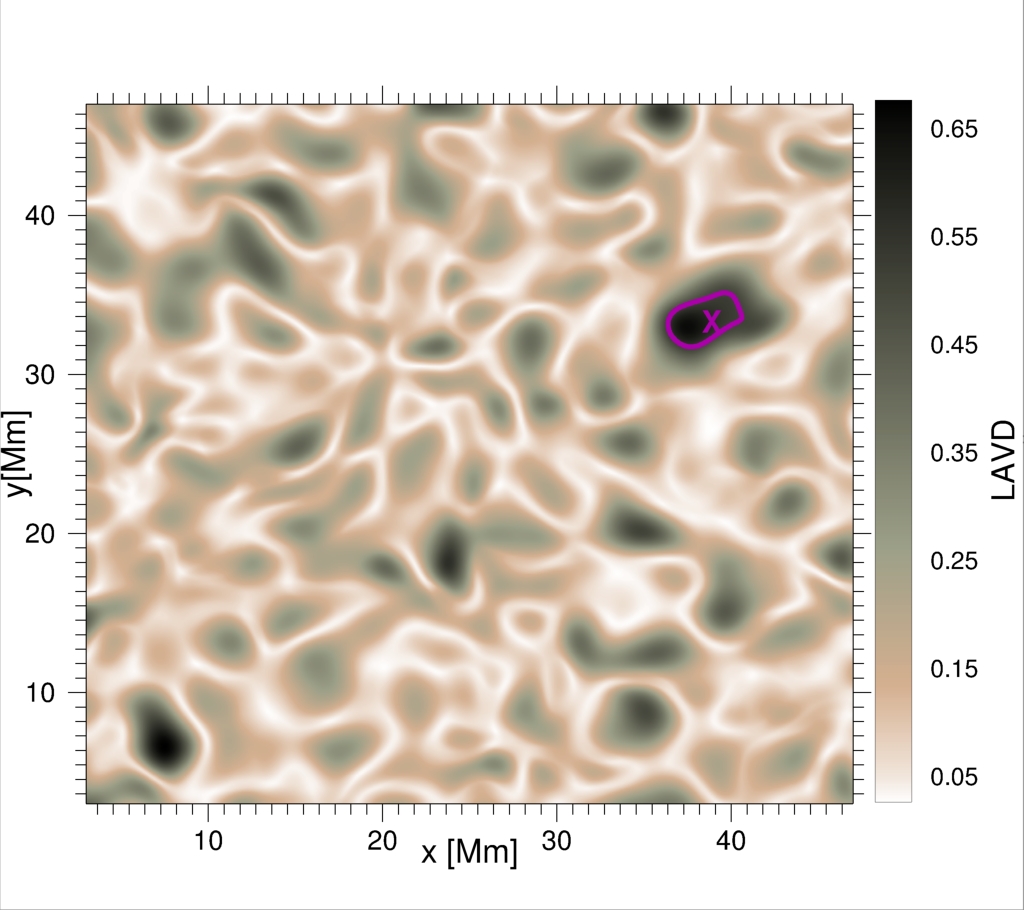}}
\qquad
\subfigure[ref1][\label{fig:6c}]{\includegraphics[trim =1mm 5mm 1mm 5mm, clip,width=0.48\textwidth]{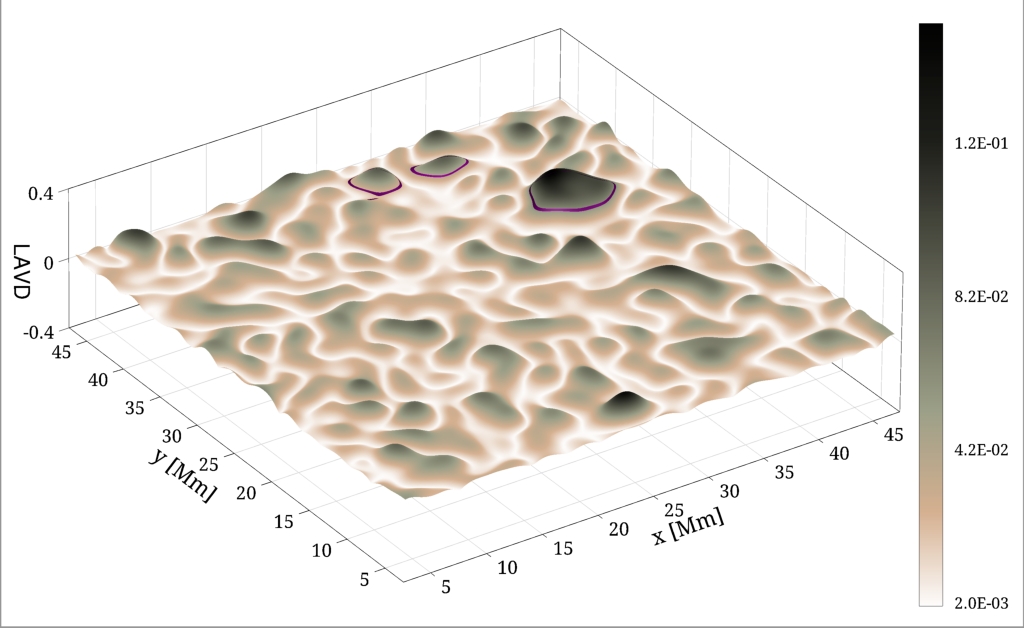}}
\qquad
\subfigure[ref1][\label{fig:6d}]{\includegraphics[trim =1mm 5mm 1mm 5mm, clip,width=0.48\textwidth]{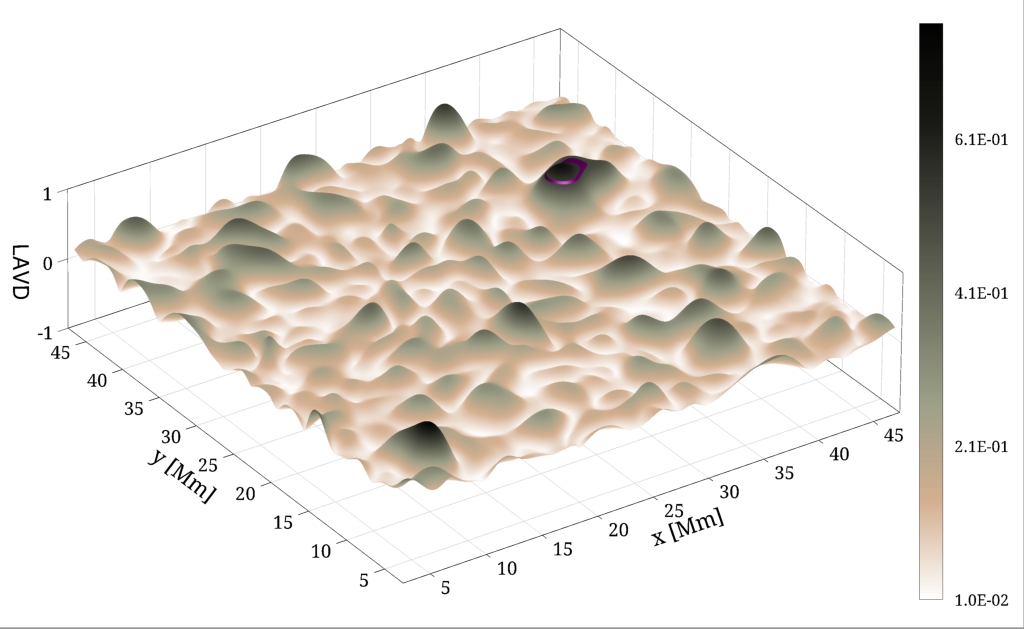}}
    \caption{Elliptic Lagrangian coherent structures of supergranular turbulence. (a) 2D plot of LAVD for the supergranular region marked by the green rectangle in Fig. \ref{fig:1}\subref{fig:1a} computed from 20:16 UT to 20:31 UT using $\tau$ = 15 min (a), and from 20:46 UT to 21:46 UT using  $\tau$ = 60 min (b). The magenta line (cross) denotes the boundary (centre) of each Lagrangian vortex. 3D surface plot of LAVD for Fig. \ref{fig:5}\subref{fig:5a} (c) and Fig. \ref{fig:5}\subref{fig:5b} (d).}
    \label{fig:6}
\end{figure*}

We considered the supergranular region marked by the green rectangle in Fig. \ref{fig:1}\subref{fig:1a} and applied the LAVD method together with the d-parameter to detect coherent objective vortices. Figures \ref{fig:6}\subref{fig:6a} and \ref{fig:6}\subref{fig:6b} show two examples of 2D plots of the LAVD field computed from 20:16 UT to 20:31 UT for $\tau$ = 15 min and from 20:46 UT to 21:46 UT for $\tau$= 60 min, respectively. It is worth mentioning that the parameter $\tau$ is not a measure of the vortex lifetime; rather, it is the finite time interval for computing LAVD.  The boundary of each Lagrangian vortex is indicated by a magenta line, denoting the outermost closed (approximately convex) line of LAVD. Figures  \ref{fig:6}\subref{fig:6c}  and  \ref{fig:6}\subref{fig:6d}  exhibit 3D surface plots of Figs.  \ref{fig:6}\subref{fig:6a}  and  \ref{fig:6}\subref{fig:6b} , respectively, presenting a 3D perspective of Lagrangian vortices detected. The centre of a Lagrangian vortex is given by the location of the centre of the spiral streamlines of the displacement vector field that obeys the d-criterion \citep{Silva_2018b}. The LAVD contour represents the material line where all the particles  experience the same intrinsic dynamic rotation \citep{Haller2016}. The large vortex areas detected by LAVD indicate that the vortical dynamics imposed by the vortices extends for a considerable length, as observed by \citet{Silva_2018b}. 

\begin{figure*}
\center
\subfigure[ref1][\label{fig:7a}]{\includegraphics[trim =1mm 10mm 1mm 10mm, clip,width=0.48\textwidth]{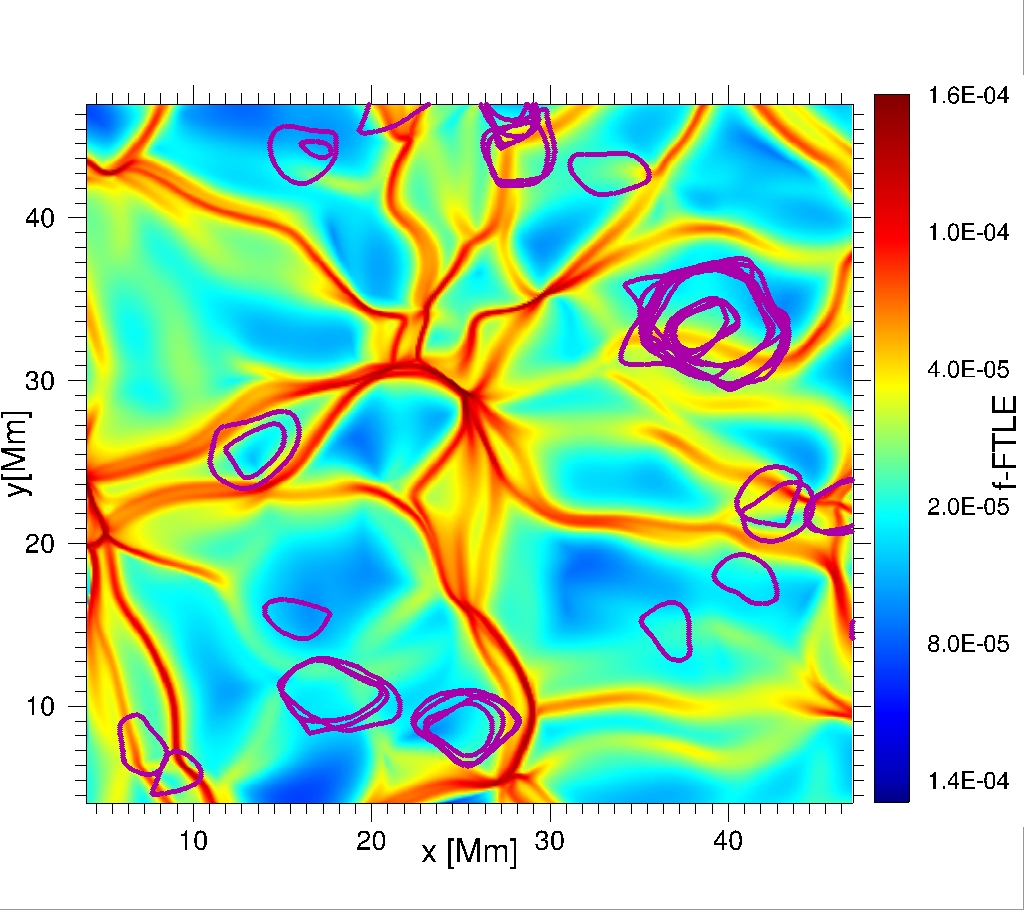}}
\qquad
\subfigure[ref1][\label{fig:7b}]{\includegraphics[trim =1mm 10mm 1mm 10mm, clip,width=0.48\textwidth]{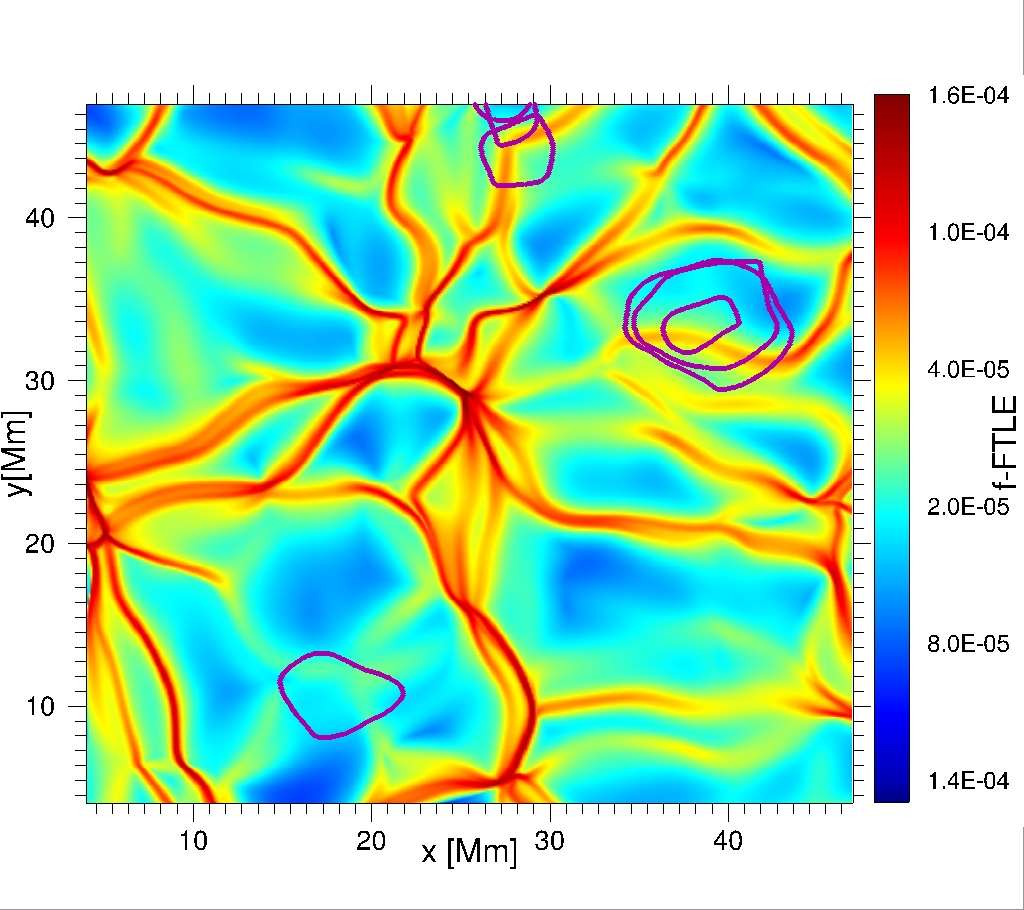}}
\qquad
\subfigure[ref1][\label{fig:7c}]{\includegraphics[trim =1mm 5mm 1mm 5mm, clip,width=0.48\textwidth]{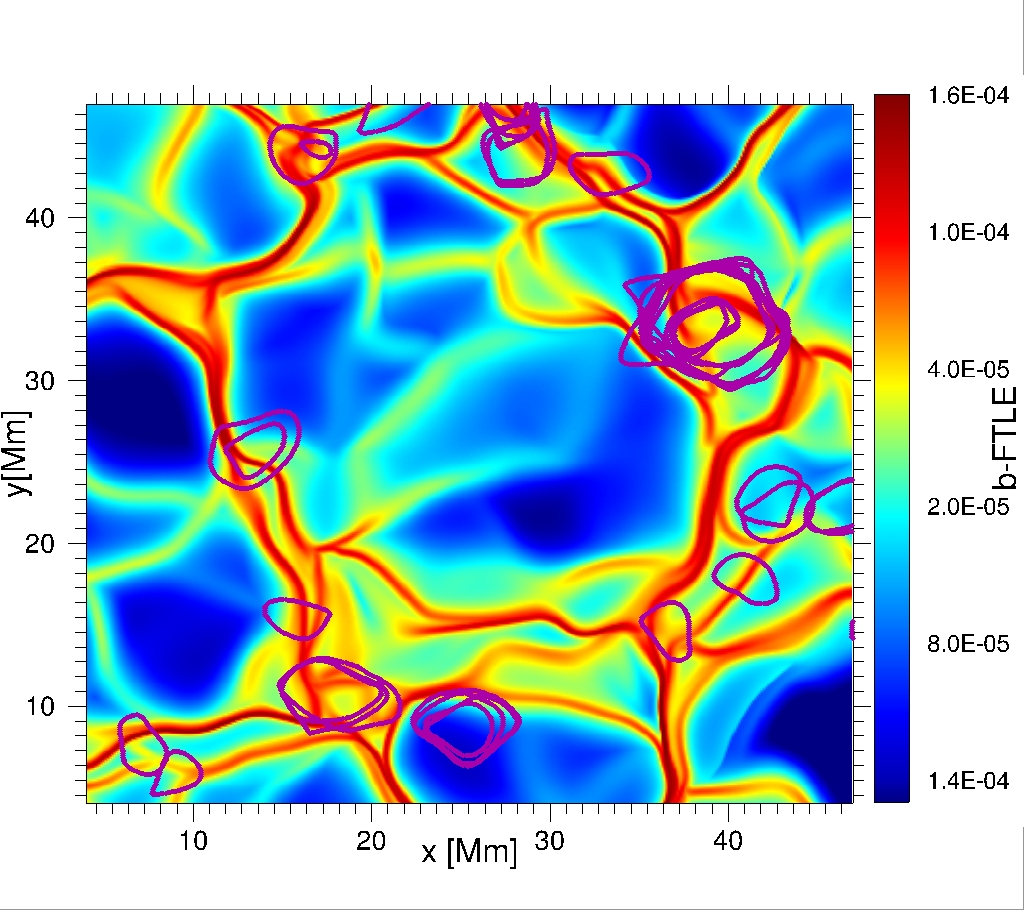}}
\qquad
\subfigure[ref1][\label{fig:7d}]{\includegraphics[trim =1mm 5mm 1mm 5mm, clip,width=0.48\textwidth]{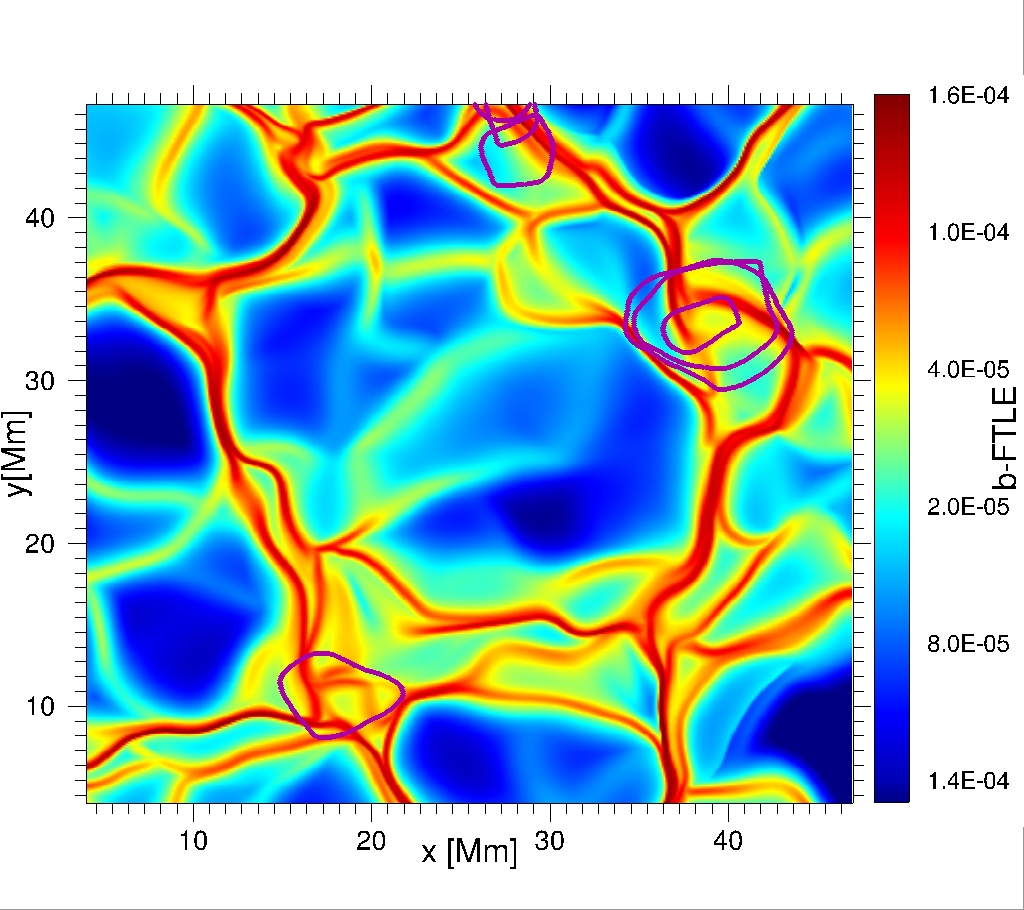}}
\caption{Superposition of elliptic and hyperbolic Lagrangian coherent structures for the time interval from 16:46:26 UT to 23:46:34 UT on 2010 November 2. Superposition of all Lagrangian vortices detected by LAVD for the time interval from 16:46:26 UT to 23:46: 34 UT, using a sliding window of $\tau$ = 15 min (a) and $\tau$ = 60 min (b), respectively, and the f-FTLE given by Fig. \ref{fig:4}\subref{fig:4a}  Superposition of all Lagrangian vortices detected by LACD for the time interval from 16:46:26 UT to 23:46:34 UT, using a sliding window of $\tau$ = 15 min (c) and $\tau$ = 60 min (d),
respectively, and the b-FTLE given by Fig. \ref{fig:4}\subref{fig:4b}. The magenta line denotes the boundary of each Lagrangian vortex.}
\label{fig:7}
\end{figure*}

\begin{figure*}
\center
\subfigure[ref1][\label{fig:8a}]{\includegraphics[trim =1mm 10mm 1mm 10mm, clip,width=0.48\textwidth]{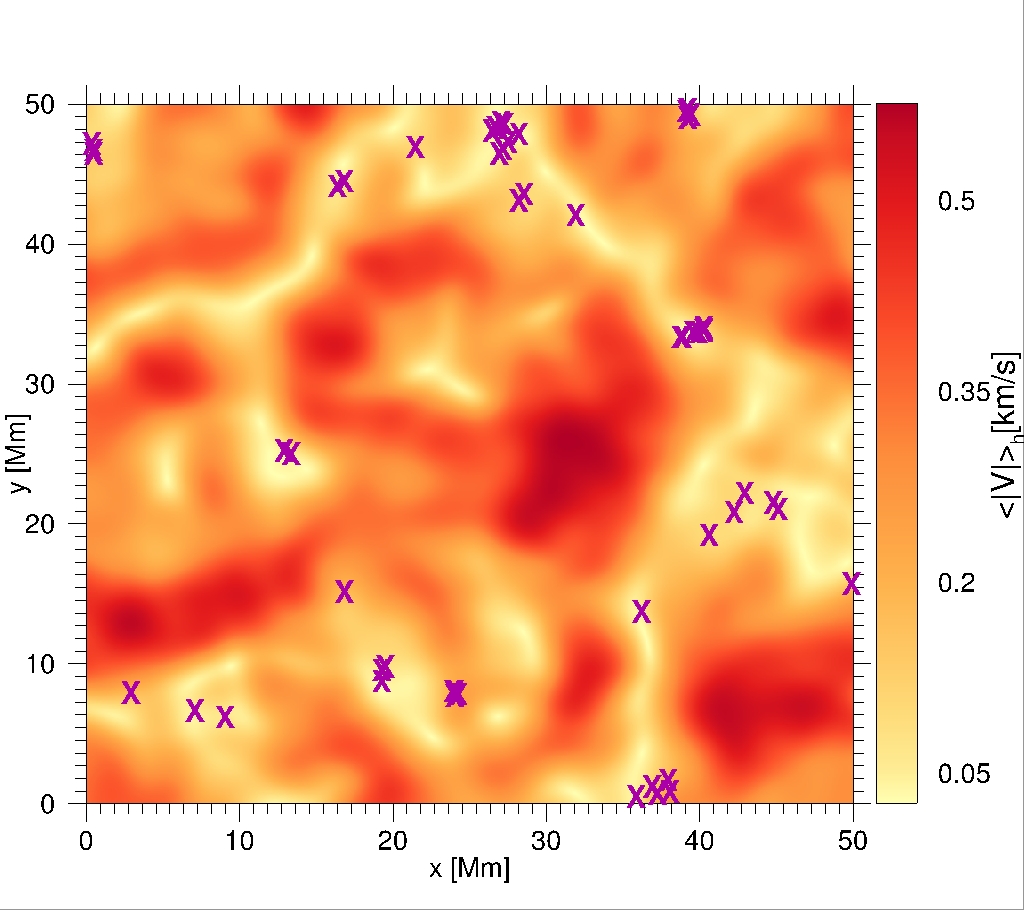}}
\qquad
\subfigure[ref1][\label{fig:8b}]{\includegraphics[trim =1mm 10mm 1mm 10mm, clip,width=0.48\textwidth]{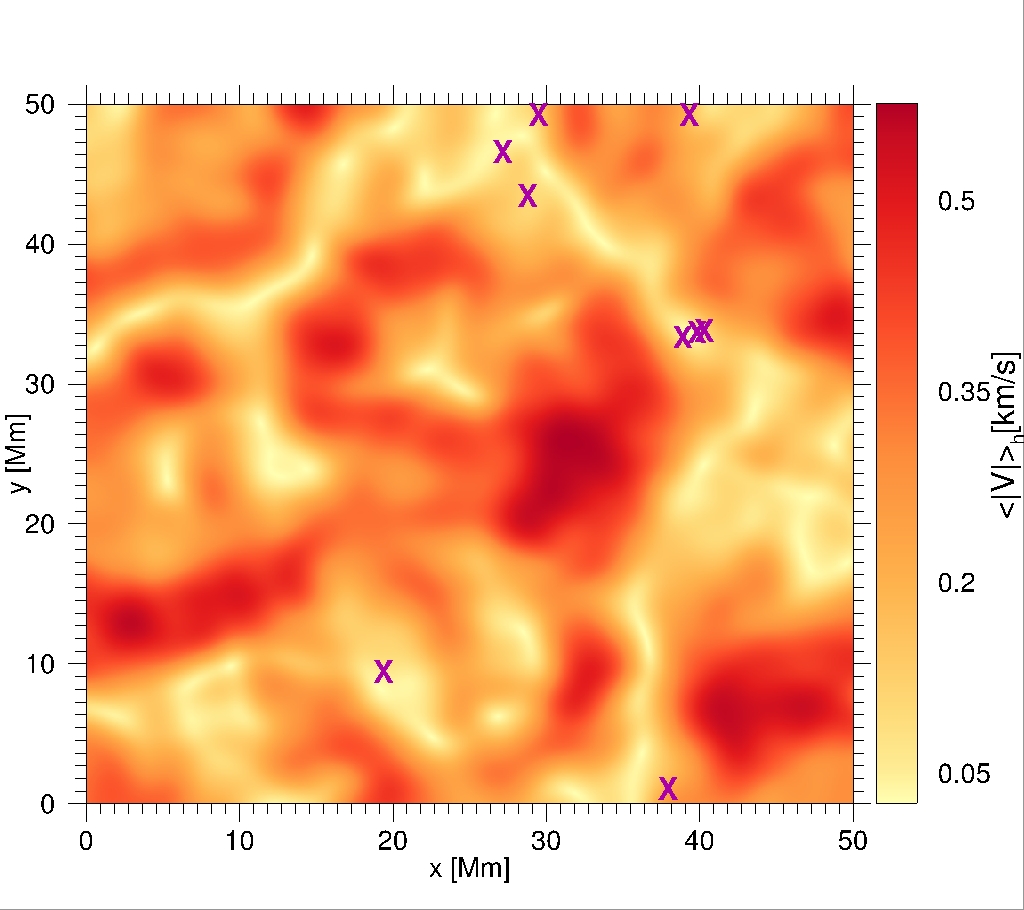}}
\qquad
\subfigure[ref1][\label{fig:8c}]{\includegraphics[trim =1mm 5mm 1mm 5mm, clip,width=0.48\textwidth]{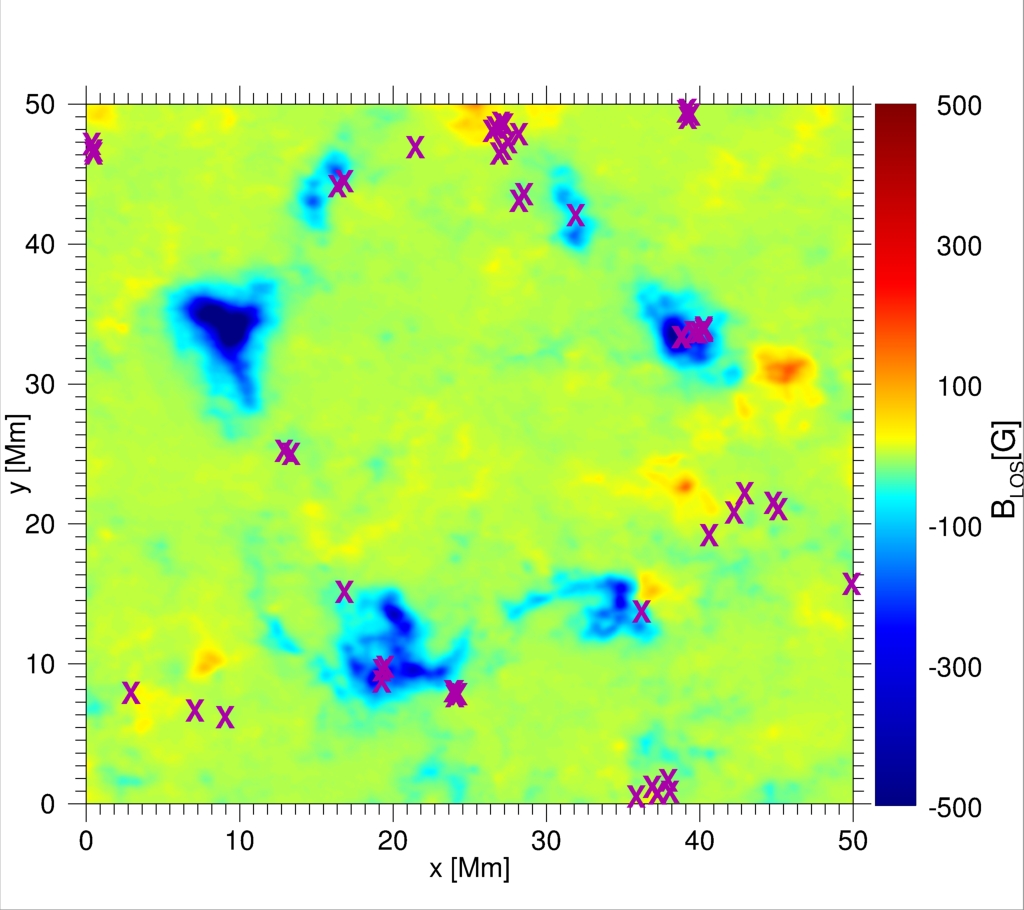}}
\qquad
\subfigure[ref1][\label{fig:8d}]{\includegraphics[trim =1mm 5mm 1mm 5mm, clip,width=0.48\textwidth]{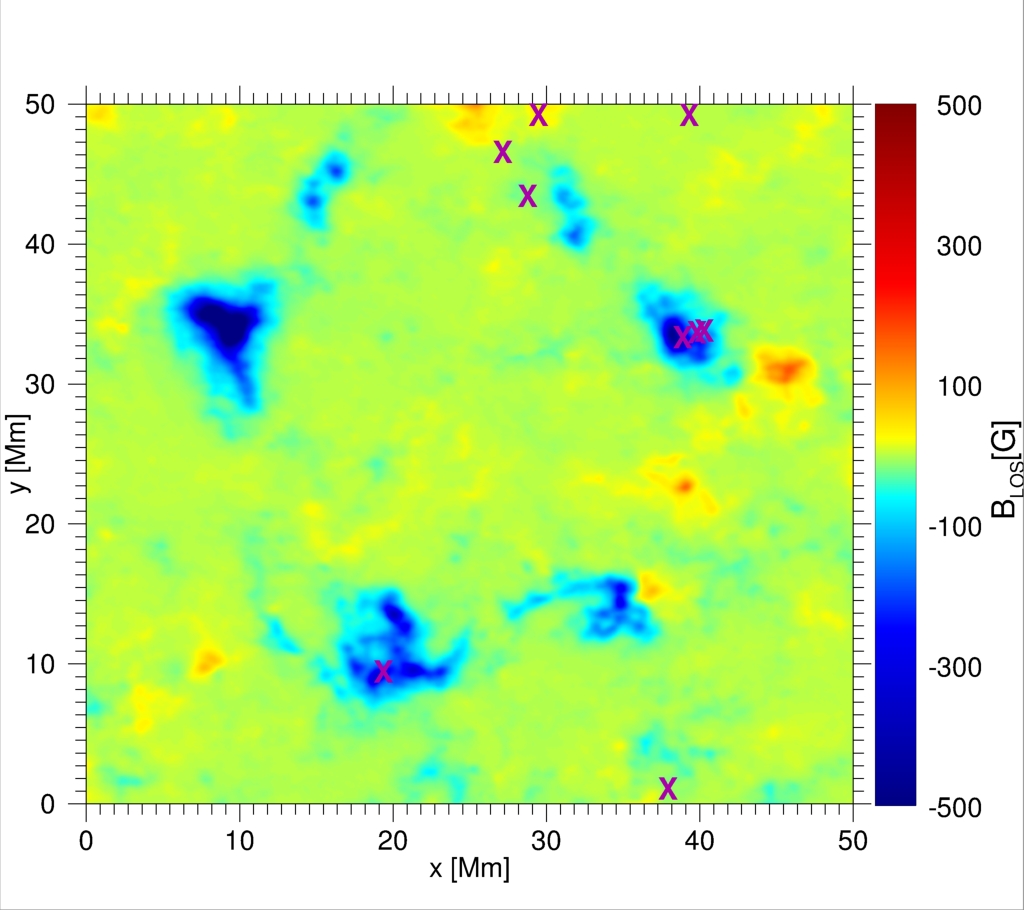}}
\caption{Superposition of Lagrangian vortex centres with the mean horizontal velocity modulus and the mean line-of-sight magnetic field for the time interval from 16:46:26 UT to 23:46:34 UT on 2010 November 2. Superposition of all Lagrangian vortex centres detected by the d-parameter for the time interval from 16:46:26 UT to 23:46:34 UT, using a sliding window of $\tau$ = 15 min (a) and $\tau$ = 60 min (b), respectively, and a background image that displays the horizontal velocity modulus time-averaged from 16:46:26 UT to 23:46:34 UT. Superposition of all Lagrangian vortex centres detected by the d-parameter
for the interval from 16:46:26 UT to 23:46:34 UT, using a sliding window of$\tau$ = 15 min (c) and$\tau$ = 60 min (d), respectively, and a background image that displays the line-of-sight magnetic field time-averaged from 16:46:26 UT to 23:46:34 UT. The magenta cross denotes the centre of each Lagrangian vortex.}
\label{fig:8}
\end{figure*}

In Figs.  \ref{fig:7}\subref{fig:7a}  and  \ref{fig:7}\subref{fig:7b}, we plot the boundary (denoted by a magenta line) of all Lagrangian vortices detected by LAVD and d-criterion using a sliding window with a duration $\tau$ during the 7-hr interval from 16:46:26 UT to 23:46:34 UT, for $\tau$ = 15 min (a) and $\tau$ = 60 min (b), respectively, superposed by the f-FTLE computed from 16:46:26 UT to 23:46:34 UT given by Fig. \ref{fig:4}\subref{fig:4a}. In Figs. \ref{fig:7}\subref{fig:7c}  and  \ref{fig:7}\subref{fig:7d}, 
the same vortex boundaries as in Figs. \ref{fig:7}\subref{fig:7a}  and  \ref{fig:7}\subref{fig:7b} are plotted superposed by the b-FTLE computed from 23:46:34 UT to 16:46:26 UT given by Fig. \ref{fig:4}\subref{fig:4b}. We can see that the elliptic LCS are distributed along the Lagrangian supergranular boundaries.
The centre (denoted by a magenta cross) of all detected Lagrangian vortices are displayed in Figs. \ref{fig:8}\subref{fig:8a}  and  \ref{fig:8}\subref{fig:8b}. Again, the vortices were detected by LAVD and d-criterion using a sliding window with a duration $\tau$ during the 7-hr interval from 16:46:26 UT to 23:46:34 UT, for $\tau$ = 15 min (a) and $\tau$ = 60 min (b), respectively, superposed by a background image that displays the horizontal velocity modulus time-averaged from 16:46:26 UT to 23:46:34 UT. In Figs. \ref{fig:8}\subref{fig:8c}  and \ref{fig:8}\subref{fig:8d}, the same Lagrangian vortex centres as in Figs. \ref{fig:8}\subref{fig:8a}  and  \ref{fig:8}\subref{fig:8b} are plotted superposed by a background image that displays the line-of-sight magnetic field time-averaged from 16:46:26 UT to 23:46:34 UT.
The total number of Lagrangian vortices detected from 16:46:26 UT to 23:46:34 UT for $\tau$ = 15 min is 50, as seen Figs. \ref{fig:8}\subref{fig:8a}  and  \ref{fig:8}\subref{fig:8c}; the total number of Lagrangian vortices detected for $\tau$ = 60 is 9, as seen in Figs. \ref{fig:8}\subref{fig:8b}  and  \ref{fig:8}\subref{fig:8d}. Note that Figs. \ref{fig:6} and \ref{fig:7} do not show all vortices detected because we have to reduce the photospheric area of Figs. \ref{fig:6} and \ref{fig:7} in view of large numerical errors in the edge region of the computed LAVD and FTLE fields, as passive particles are advected toward the domain boundaries.

\section{ DISCUSSIONS AND CONCLUSION}

Our present study confirmed in Fig. \ref{fig:1}\subref{fig:1a} that a persistent vortex is located at a supergranular vertex where flows from several nearby supergranular cells collide and interact, leading to the formation of a persistent vortex associated with photospheric swirling flows, seen in Fig. \ref{fig:1}\subref{fig:1b}. The mean Dopplergram obtained by averaging over a 7-hr duration of the high-resolution Hinode dataset shows in Fig. \ref{fig:1}\subref{fig:1a}  that the interior of a supergranular cell is dominated by upflows, whereas the supergranular boundaries are dominated by downflows; the horizontal velocities driven by turbulent magnetoconvection are directed radially outward from the mean centre of a supergranular cell, given by the local minimum of the mean horizontal velocity field, toward its boundaries. 

Our LCS analysis is based on the 2-hourly time-averaged horizontal velocity field deduced by LCT from Hinode/NFI images of the line-of-sight magnetic field, exemplified by Fig. \ref{fig:2} for the initial and final frames of a 7-hr continuous Hinode observation of the disk centre of a quiet Sun. The Eulerian magnetograms of supergranular cells given in Fig. \ref{fig:3} indicates that the IN magnetic elements undergo a variety of dynamical processes as they are transported outward from the supergranular centre to its boundary. The Eulerian supergranular boundaries shown in Fig. \ref{fig:3} separate the IN and  {NE} magnetic elements, where IN elements interact with NE elements and deposit their flux in the  {NE}. We found a strong correlation between the spatial distribution of magnetic flux and hyperbolic LCS as shown in Fig. \ref{fig:5}.
The repelling LCS, obtained by computing the f-FTLE for a duration of 7 hr, shown in Fig. \ref{fig:4}, identifies the locally strongest repelling material lines which exert the most impact on the diverging transport of supergranular flows.  The Lagrangian center of the supergranule can be defined as the local maximum of the f-FTLE field, as it marks the region where particles diverge the most during the considered time interval. We demonstrated that the interior of supergranules is dominated by repelling material lines which describe spreading motions as in \cite{Giannattasio2013}.  
The presence of such lines also creates transport barriers within IN regions that prevent mixing of IN magnetic patches across them.  The Lagrangian centres of neighboring supergranular cells are interconnected by ridges of the repelling LCS, as seen in Fig. \ref{fig:4}. As the magnetic patches diverge from the repelling lines, they tend to create magnetic voids in the regions where they cross the attracting hyperbolic LCS. 
Since the f-FTLE is closely linked to the squashing Q-factor \citep{Demoulin1996, Yeates2012, Chian2014, Inoue2016}, the repelling LCS also identify the most likely sites for the occurrence of magnetic reconnection involving element-element interactions such as the cancellation events seen in \textit{Hinode}/NFI magnetograms. The attracting LCS obtained by computing the b-FTLE for a duration of 7 hr, shown in Fig. \ref{fig:4}, display thin ridges of large positive b-FTLE, representing the locally strongest attracting material lines which exert the most impact on the converging transport of supergranular flows. Therefore, they act as the Lagrangian supergranular boundaries which identify the preferential sites of photospheric downdrafts; the attracting LCS are co-spatial with the network of high magnetic flux concentration, which confirms the trapping of magnetic elements in the NE region observed by \cite{Orozco2012}. The attracting material lines provide the transport barriers between the supergranules,  indicating that the mixing of IN elements is restrained to the NE region of supergranule as demonstrated by \cite{Gosic2014}, who applied an automatic feature tracking algorithm to follow the evolution of IN and NE magnetic elements. They showed that the IN magnetic elements continually transfer their flux to the NE by interacting with the NE magnetic elements via the merging (cancellation) processes that add (remove) flux. Our results are also in accordance with \cite{Agrawal2018}; they applied the concepts of complex systems with high cadence MURaM simulations and found a strong tendency of a supergranular vertex to host magnetic features and also a longer correlation time for NE elements. In contrast to the simple picture of supergranular boundaries given by the Eulerian magnetograms in Fig. \ref{fig:3}, we see in Fig. \ref{fig:7} that the Lagrangian supergranular boundaries are very complex, resulting from the  {nonlinear} interactions between supergranular cells. In Fig. \ref{fig:5} we plot the superposition of the hyperbolic repelling (a) and attracting (b) LCS with the time-averaged magnetogram of the line-of-sight magnetic field from 16:46:26 UT to 23:46:34 UT. Figure \ref{fig:5a} suggests that the IN magnetic elements advected by turbulent convective flows diverge preferentially along the repelling LCS until they reach the Lagrangian boundaries of supergranular cells, then converge preferentially to the attracting LCS \ref{fig:5b}, wherein they deposit their magnetic fluxes, making possible the concentration and amplification of strong NE magnetic fields.

By superposing the boundary of all Lagrangian vortices detected during 7 hr with the  {b-FTLE} in Fig. \ref{fig:7}, we see that the Lagrangian vortices are mostly located in the high-value regions of the  {b-FTLE}, where the ridges of the attracting LCS provide the sinks for photospheric flows. Therefore, most of those vortices are very likely to be a consequence of the bathtub effect. 
Physically, the contours obtained for the vortices mean that for the time interval considered the fluid elements within the contour are trapped in that region. Therefore, magnetic elements patches tend to be trapped by the vortices which cause the magnetic field intensity to increase. Once the vortex ceases to exist, the magnetic elements will be free to go to other regions, leading to magnetic flux decay as observed by \cite{Requerey2018}.  By superposing the centre of all Lagrangian vortices detected with the 7-hr time-averaged horizontal velocity field and line-of-sight magnetic field in Fig. \ref{fig:8}, we confirm that the supergranular junctions are the preferential sites for the concentration of intense magnetic flux as well as the formation of Lagrangian vortices and associated kG magnetic flux tubes. In particular, Figs.  \ref{fig:6}-\ref{fig:8} confirm the detection of the persistent vortex that appears n the black rectangle region of Fig. \ref{fig:1a}.

The study of supergranulation is relevant for understanding solar dynamo, coronal heating and eruption. LCS provide a powerful tool to determine the transport properties of photospheric flows and the complex dynamics of solar magnetic fields \citep{Aschwanden2018, Rincon2018, Roudier2018}. Our analysis revealed that the supergranular turbulence is governed by hyperbolic and elliptic LCS, demonstrating that supergranulation-scale convection plays a key role in the nonlinear dynamics of local and global solar magnetic fields at the interface between solar interior and solar atmosphere. The conclusions of this paper are based on a 7-hr LCS data analysis for a photospheric region of 50 Mm x 50 Mm. Future works can apply this technique to different temporal and spatial scales of quiet- and active-Sun. In conclusion, the detection of Lagrangian coherent structures in supergranular turbulence can deepen our understanding of the nonlinear  dynamics and complex structures of the photosphere, as well as improve our ability to monitor and predict solar and astrophysical eruptive events such as flares and coronal mass ejections.

\section*{Acknowledgements}
This work is dedicated to the 160th anniversary of the first report by Richard Christopher Carrington (a graduate of Trinity College, Cambridge) of a solar superflare on 1859 September 1 and its relation to a major geomagnetic storm. The data used here were acquired in the framework of the \textit{Hinode} Operation Plan 151 {\em ``Flux replacement in the solar network and internetwork.''} We thank the \textit{Hinode} Chief Observers for the efforts they made to accommodate our demanding observations. \textit{Hinode} is a Japanese mission developed and launched by ISAS/JAXA, with NAOJ as a domestic partner and NASA and STFC (UK) as international partners. It is operated by these agencies in co-operation with ESA and NSC (Norway). SSAS acknowledges financial support from  agency CAPES (88882.316962/2019-01, Brazil). ELR acknowledges financial support from CNPq (Brazil), partial financial support from CAPES (Brazil) and from FAPESP (Brazil).


\begin{thebibliography}{}
\makeatletter
\relax
\def\mn@urlcharsother{\let\do\@makeother \do\$\do\&\do\#\do\^\do\_\do\%\do\~}
\def\mn@doi{\begingroup\mn@urlcharsother \@ifnextchar [ {\mn@doi@}
  {\mn@doi@[]}}
\def\mn@doi@[#1]#2{\def\@tempa{#1}\ifx\@tempa\@empty \href
  {http://dx.doi.org/#2} {doi:#2}\else \href {http://dx.doi.org/#2} {#1}\fi
  \endgroup}
\def\mn@eprint#1#2{\mn@eprint@#1:#2::\@nil}
\def\mn@eprint@arXiv#1{\href {http://arxiv.org/abs/#1} {{\tt arXiv:#1}}}
\def\mn@eprint@dblp#1{\href {http://dblp.uni-trier.de/rec/bibtex/#1.xml}
  {dblp:#1}}
\def\mn@eprint@#1:#2:#3:#4\@nil{\def\@tempa {#1}\def\@tempb {#2}\def\@tempc
  {#3}\ifx \@tempc \@empty \let \@tempc \@tempb \let \@tempb \@tempa \fi \ifx
  \@tempb \@empty \def\@tempb {arXiv}\fi \@ifundefined
  {mn@eprint@\@tempb}{\@tempb:\@tempc}{\expandafter \expandafter \csname
  mn@eprint@\@tempb\endcsname \expandafter{\@tempc}}}

\bibitem[\protect\citeauthoryear{{Agrawal}, {Rast}, {Go{\v{s}}i{\'c}}, {Bellot
  Rubio}  \& {Rempel}}{{Agrawal} et~al.}{2018}]{Agrawal2018}
{Agrawal} P.,  {Rast} M.~P.,  {Go{\v{s}}i{\'c}} M.,  {Bellot Rubio} L.~R.,
  {Rempel} M.,  2018, \mn@doi [\apj] {10.3847/1538-4357/aaa251}, \href
  {https://ui.adsabs.harvard.edu/\#abs/2018ApJ...854..118A} {854, 118}

\bibitem[\protect\citeauthoryear{{Aschwanden}, {Go{\v{s}}ic}, {Hurlburt}  \&
  {Scullion}}{{Aschwanden} et~al.}{2018}]{Aschwanden2018}
{Aschwanden} M.~J.,  {Go{\v{s}}ic} M.,  {Hurlburt} N.~E.,   {Scullion} E.,
  2018, \mn@doi [\apj] {10.3847/1538-4357/aae08b}, \href
  {https://ui.adsabs.harvard.edu/\#abs/2018ApJ...866...73A} {866, 73}

\bibitem[\protect\citeauthoryear{{Attie}, {Innes}  \& {Potts}}{{Attie}
  et~al.}{2009}]{Attie2009}
{Attie} R.,  {Innes} D.~E.,   {Potts} H.~E.,  2009, \mn@doi [\aap]
  {10.1051/0004-6361:200811258}, \href
  {https://ui.adsabs.harvard.edu/\#abs/2009A&A...493L..13A} {493, L13}

\bibitem[\protect\citeauthoryear{Bellot~Rubio \&
  Orozco~Su{\'a}rez}{Bellot~Rubio \& Orozco~Su{\'a}rez}{2019}]{Bellot2019}
Bellot~Rubio L.,  Orozco~Su{\'a}rez D.,  2019, \mn@doi [Living Rev. Sol. Phys.] {10.1007/s41116-018-0017-1}, 16, 1

\bibitem[\protect\citeauthoryear{Beron-Vera, Olascoaga, Wang, Tri{\~n}anes  \&
  P{\'e}rez-Brunius}{Beron-Vera et~al.}{2018}]{Beron-Vera2018}
Beron-Vera F.~J.,  Olascoaga M.~J.,  Wang Y.,  Tri{\~n}anes J.,
  P{\'e}rez-Brunius P.,  2018, \mn@doi [Sci Rep] {10.1038/s41598-018-29582-5},
  8, 11275

\bibitem[\protect\citeauthoryear{{Borgogno}, {Grasso}, {Pegoraro}  \&
  {Schep}}{{Borgogno} et~al.}{2011}]{Borgogno2011}
{Borgogno} D.,  {Grasso} D.,  {Pegoraro} F.,   {Schep} T.~J.,  2011, \mn@doi
  [Phys. Plasmas] {10.1063/1.3647339}, \href
  {https://ui.adsabs.harvard.edu/\#abs/2011PhPl...18j2307B} {18, 102307}

\bibitem[\protect\citeauthoryear{{Brandt}, {Scharmert}, {Ferguson}, {Shine},
  {Tarbell}  \& {Title}}{{Brandt} et~al.}{1988}]{Brandt1988}
{Brandt} P.~N.,  {Scharmert} G.~B.,  {Ferguson} S.,  {Shine} R.~A.,  {Tarbell}
  T.~D.,   {Title} A.~M.,  1988, \mn@doi [\nat] {10.1038/335238a0}, \href
  {http://adsabs.harvard.edu/abs/1988Natur.335..238B} {335, 238}

\bibitem[\protect\citeauthoryear{Caroli, Giannattasio, Fanfoni, Del~Moro,
  Consolini  \& Berrilli}{Caroli et~al.}{2015}]{Caroli2015}
Caroli A.,  Giannattasio F.,  Fanfoni M.,  Del~Moro D.,  Consolini G.,
  Berrilli F.,  2015, \mn@doi [J. Plasma Phys.]
  {10.1017/S0022377815000872}, 81, 495810514

\bibitem[\protect\citeauthoryear{{Carrington}}{{Carrington}}{1859}]{Carrington1859}
{Carrington} R.~C.,  1859, \mn@doi [\mnras] {10.1093/mnras/20.1.13}, \href
  {http://adsabs.harvard.edu/abs/1859MNRAS..20...13C} {20, 13}

\bibitem[\protect\citeauthoryear{{Chian}, {Rempel}, {Aulanier}, {Schmieder},
  {Shadden}, {Welsch}  \& {Yeates}}{{Chian} et~al.}{2014}]{Chian2014}
{Chian} A. C.~L.,  {Rempel} E.~L.,  {Aulanier} G.,  {Schmieder} B.,  {Shadden}
  S.~C.,  {Welsch} B.~T.,   {Yeates} A.~R.,  2014, \mn@doi [\apj]
  {10.1088/0004-637X/786/1/51}, \href
  {https://ui.adsabs.harvard.edu/\#abs/2014ApJ...786...51C} {786, 51}

\bibitem[\protect\citeauthoryear{{D\'emoulin}, {Henoux}, {Priest}  \& {Mand
  rini}}{{D\'emoulin} et~al.}{1996}]{Demoulin1996}
{D\'emoulin} P.,  {Henoux} J.~C.,  {Priest} E.~R.,   {Mand rini} C.~H.,  1996,
  \aap, \href {https://ui.adsabs.harvard.edu/\#abs/1996A&A...308..643D} {308,
  643}

\bibitem[\protect\citeauthoryear{{Giannattasio}, {Del Moro}, {Berrilli},
  {Bellot Rubio}, {Go{\v s}i{\'c}}  \& {Orozco Su{\'a}rez}}{{Giannattasio}
  et~al.}{2013}]{Giannattasio2013}
{Giannattasio} F.,  {Del Moro} D.,  {Berrilli} F.,  {Bellot Rubio} L.,  {Go{\v
  s}i{\'c}} M.,   {Orozco Su{\'a}rez} D.,  2013, \mn@doi [\apjl]
  {10.1088/2041-8205/770/2/L36}, \href
  {http://adsabs.harvard.edu/abs/2013ApJ...770L..36G} {770, L36}

\bibitem[\protect\citeauthoryear{{Giannattasio}, {Berrilli}, {Biferale}, {Del
  Moro}, {Sbragaglia}, {Bellot Rubio}, {Go{\v{s}}i{\'c}}  \& {Orozco
  Su{\'a}rez}}{{Giannattasio} et~al.}{2014a}]{Giannattasio2014a}
{Giannattasio} F.,  {Berrilli} F.,  {Biferale} L.,  {Del Moro} D.,
  {Sbragaglia} M.,  {Bellot Rubio} L.,  {Go{\v{s}}i{\'c}} M.,   {Orozco
  Su{\'a}rez} D.,  2014a, \mn@doi [\aap] {10.1051/0004-6361/201424380}, \href
  {https://ui.adsabs.harvard.edu/\#abs/2014A&A...569A.121G} {569, A121}

\bibitem[\protect\citeauthoryear{{Giannattasio}, {Stangalini}, {Berrilli}, {Del
  Moro}  \& {Bellot Rubio}}{{Giannattasio} et~al.}{2014b}]{Giannattasio2014b}
{Giannattasio} F.,  {Stangalini} M.,  {Berrilli} F.,  {Del Moro} D.,   {Bellot
  Rubio} L.,  2014b, \mn@doi [\apj] {10.1088/0004-637X/788/2/137}, \href
  {https://ui.adsabs.harvard.edu/\#abs/2014ApJ...788..137G} {788, 137}

\bibitem[\protect\citeauthoryear{{Giannattasio}, {Berrilli}, {Consolini}, {Del
  Moro}, {Go{\v{s}}i{\'c}}  \& {Bellot Rubio}}{{Giannattasio}
  et~al.}{2018}]{Giannattasio2018}
{Giannattasio} F.,  {Berrilli} F.,  {Consolini} G.,  {Del Moro} D.,
  {Go{\v{s}}i{\'c}} M.,   {Bellot Rubio} L.,  2018, \mn@doi [\aap]
  {10.1051/0004-6361/201730583}, \href
  {https://ui.adsabs.harvard.edu/\#abs/2018A&A...611A..56G} {611, A56}

\bibitem[\protect\citeauthoryear{Gold \& Hoyle}{Gold \& Hoyle}{1960}]{Gold1960}
Gold T.,  Hoyle F.,  1960, \mn@doi [\mnras] {10.1093/mnras/120.2.89}, \href
  {http://adsabs.harvard.edu/abs/1960MNRAS.120...89G} {120, 89}

\bibitem[\protect\citeauthoryear{Go{\v{s}}i{\'{c}}, {Bellot Rubio},
  Su{\'{a}}rez, Katsukawa  \& del Toro~Iniesta}{Go{\v{s}}i{\'{c}}
  et~al.}{2014}]{Gosic2014}
Go{\v{s}}i{\'{c}} M.,  {Bellot Rubio} L.~R.,  Su{\'{a}}rez D.~O.,  Katsukawa
  Y.,   del Toro~Iniesta J.~C.,  2014, \mn@doi [\apj]
  {10.1088/0004-637x/797/1/49}, 797, 49

\bibitem[\protect\citeauthoryear{Go{\v{s}}i{\'{c}}, {Bellot Rubio}, del
  Toro~Iniesta, Su{\'{a}}rez  \& Katsukawa}{Go{\v{s}}i{\'{c}}
  et~al.}{2016}]{Gosic2016}
Go{\v{s}}i{\'{c}} M.,  {Bellot Rubio} L.~R.,  del Toro~Iniesta J.~C.,
  Su{\'{a}}rez D.~O.,   Katsukawa Y.,  2016, \mn@doi [\apjl]
  {10.3847/0004-637x/820/1/35}, 820, 35

\bibitem[\protect\citeauthoryear{Hadjighasem, Farazmand, Blazevski, Froyland
  \& Haller}{Hadjighasem et~al.}{2017}]{Hadjighasem2017}
Hadjighasem A.,  Farazmand M.,  Blazevski D.,  Froyland G.,   Haller G.,  2017,
  \mn@doi [Chaos]
  {10.1063/1.4982720}, 27, 053104

\bibitem[\protect\citeauthoryear{Haller}{Haller}{2015}]{Haller2015}
Haller G.,  2015, \mn@doi [Annu. Rev. Fluid. Mech.]
  {10.1146/annurev-fluid-010313-141322}, 47, 137

\bibitem[\protect\citeauthoryear{Haller \& Yuan}{Haller \&
  Yuan}{2000}]{Haller2000}
Haller G.,  Yuan G.,  2000, \mn@doi [Physica D]
  {10.1016/S0167-2789(00)00142-1}, 147, 352

\bibitem[\protect\citeauthoryear{Haller, Hadjighasem, Farazmand  \&
  Huhn}{Haller et~al.}{2016}]{Haller2016}
Haller G.,  Hadjighasem A.,  Farazmand M.,   Huhn F.,  2016, \mn@doi [J. Fluid Mech.] {10.1017/jfm.2016.151}, 795, 136

\bibitem[\protect\citeauthoryear{Hotta, Iijima  \& Kusano}{Hotta
  et~al.}{2019}]{Hotta2019}
Hotta H.,  Iijima H.,   Kusano K.,  2019, \mn@doi [ Sci. Adv.]
  {10.1126/sciadv.aau2307}, 5: eaau2307

\bibitem[\protect\citeauthoryear{{Howard}}{{Howard}}{1971}]{Howard1971}
{Howard} R.,  1971, \mn@doi [\solphys] {10.1007/BF00154497}, \href
  {https://ui.adsabs.harvard.edu/\#abs/1971SoPh...16...21H} {16, 21}

\bibitem[\protect\citeauthoryear{Iida, Yokoyama  \& Ichimoto}{Iida
  et~al.}{2010}]{Iida2010}
Iida Y.,  Yokoyama T.,   Ichimoto K.,  2010, \mn@doi [\apj]
  {10.1088/0004-637x/713/1/325}, 713, 325

\bibitem[\protect\citeauthoryear{Iida, Hagenaar  \& Yokoyama}{Iida
  et~al.}{2012}]{Iida2012}
Iida Y.,  Hagenaar H.~J.,   Yokoyama T.,  2012, \mn@doi [\apj]
  {10.1088/0004-637x/752/2/149}, 752, 149

\bibitem[\protect\citeauthoryear{Inoue, Hayashi  \& Kusano}{Inoue
  et~al.}{2016}]{Inoue2016}
Inoue S.,  Hayashi K.,   Kusano K.,  2016, \mn@doi [\apj]
  {10.3847/0004-637x/818/2/168}, 818, 168

\bibitem[\protect\citeauthoryear{{Kamide} \& {Chian}}{{Kamide} \&
  {Chian}}{2007}]{Kamide2007}
{Kamide} Y.,  {Chian} A.~C.-L.,  2007, {Handbook of the Solar-Terrestrial
  Environment}.
Springer, Berlin.

\bibitem[\protect\citeauthoryear{{Liu}, {Zhang}, {Ai}, {Wang}  \&
  {Zirin}}{{Liu} et~al.}{1994}]{Liu1994}
{Liu} Y.,  {Zhang} H.,  {Ai} G.,  {Wang} H.,   {Zirin} H.,  1994, \aap, \href
  {http://adsabs.harvard.edu/abs/1994A\%26A...283..215L} {283, 215}

\bibitem[\protect\citeauthoryear{Mathur, Haller, Peacock, Ruppert-Felsot  \&
  Swinney}{Mathur et~al.}{2007}]{Mathur2007}
Mathur M.,  Haller G.,  Peacock T.,  Ruppert-Felsot J.~E.,   Swinney H.~L.,
  2007, \mn@doi [Phys. Rev. Lett.] {10.1103/PhysRevLett.98.144502}, 98, 144502

\bibitem[\protect\citeauthoryear{{Nordlund}}{{Nordlund}}{1985}]{Norlund1985}
{Nordlund} A.,  1985, \mn@doi [\solphys] {10.1007/BF00158429}, \href
  {https://ui.adsabs.harvard.edu/\#abs/1985SoPh..100..209N} {100, 209}

\bibitem[\protect\citeauthoryear{{November} \& {Simon}}{{November} \&
  {Simon}}{1988}]{November1988}
{November} L.~J.,  {Simon} G.~W.,  1988, \mn@doi [\apj] {10.1086/166758}, \href
  {http://adsabs.harvard.edu/abs/1988ApJ...333..427N} {333, 427}

\bibitem[\protect\citeauthoryear{{Orozco Su{\'a}rez} et~al.,}{{Orozco
  Su{\'a}rez} et~al.}{2007a}]{Orozco2007a}
{Orozco Su{\'a}rez} D.,  et~al., 2007a, \mn@doi [\pasj]
  {10.1093/pasj/59.sp3.S837}, 59, S837

\bibitem[\protect\citeauthoryear{{Orozco Su{\'a}rez} et~al.,}{{Orozco
  Su{\'a}rez} et~al.}{2007b}]{Orozco2007b}
{Orozco Su{\'a}rez} D.,  et~al., 2007b, \mn@doi [\apjl] {10.1086/524139}, 670,
  L61

\bibitem[\protect\citeauthoryear{{Orozco Su{\'a}rez}, {Katsukawa}  \& {Bellot
  Rubio}}{{Orozco Su{\'a}rez} et~al.}{2012}]{Orozco2012}
{Orozco Su{\'a}rez} D.,  {Katsukawa} Y.,   {Bellot Rubio} L.~R.,  2012, \mn@doi
  [\apjl] {10.1088/2041-8205/758/2/L38}, \href
  {http://adsabs.harvard.edu/abs/2012ApJ...758L..38O} {758, L38}

\bibitem[\protect\citeauthoryear{{Padberg}, {Hauff}, {Jenko}  \&
  {Junge}}{{Padberg} et~al.}{2007}]{Padberg2007}
{Padberg} K.,  {Hauff} T.,  {Jenko} F.,   {Junge} O.,  2007, \mn@doi [New J. Phys.] {10.1088/1367-2630/9/11/400}, \href
  {https://ui.adsabs.harvard.edu/\#abs/2007NJPh....9..400P} {9, 400}

\bibitem[\protect\citeauthoryear{{Parker}}{{Parker}}{1955}]{Parker1955}
{Parker} E.~N.,  1955, \mn@doi [\apj] {10.1086/146087}, \href
  {http://adsabs.harvard.edu/abs/1955ApJ...122..293P} {122, 293}

\bibitem[\protect\citeauthoryear{Parker}{Parker}{1957}]{Parker1957}
Parker E.~N.,  1957, \mn@doi [\jgr] {10.1029/JZ062i004p00509}, \href
  {http://adsabs.harvard.edu/abs/1957JGR....62..509P} {62, 509}

\bibitem[\protect\citeauthoryear{Rempel, Chian  \& Brandenburg}{Rempel
  et~al.}{2011}]{Rempel2011}
Rempel E.~L.,  Chian A. C.-L.,   Brandenburg A.,  2011, \mn@doi [\apj]
  {10.1088/2041-8205/735/1/l9}, 735, L9

\bibitem[\protect\citeauthoryear{Rempel, Chian  \& Brandenburg}{Rempel
  et~al.}{2012}]{Rempel2012}
Rempel E.~L.,  Chian A. C.-L.,   Brandenburg A.,  2012, \mn@doi [Physica
  Scripta] {10.1088/0031-8949/86/01/018405}, 86, 018405

\bibitem[\protect\citeauthoryear{Rempel, Chian, Brandenburg, Mu\~noz  \&
  Shadden}{Rempel et~al.}{2013}]{Rempel2013}
Rempel E.~L.,  Chian A. C.-L.,  Brandenburg A.,  Mu\~noz P.~R.,   Shadden
  S.~C.,  2013, \mn@doi [J. Fluid Mech.] {10.1017/jfm.2013.290},
  729, 309

\bibitem[\protect\citeauthoryear{Rempel, Chian, Beron-Vera, Haller  \&
  Szanyi}{Rempel et~al.}{2017}]{Rempel2017}
Rempel E.~L.,  Chian A. C.-L.,  Beron-Vera F.~J.,  Haller G.,   Szanyi S.,
  2017, \mn@doi [MNRAS Letters] {10.1093/mnrasl/slw248}, 466, L108

\bibitem[\protect\citeauthoryear{{Requerey}, {Cobo}, {Go{\v{s}}i{\'c}}  \&
  {Bellot Rubio}}{{Requerey} et~al.}{2018}]{Requerey2018}
{Requerey} I.~S.,  {Cobo} B.~R.,  {Go{\v{s}}i{\'c}} M.,   {Bellot Rubio} L.~R.,
   2018, \mn@doi [\aap] {10.1051/0004-6361/201731842}, \href
  {https://ui.adsabs.harvard.edu/\#abs/2018A&A...610A..84R} {610, A84}

\bibitem[\protect\citeauthoryear{Rincon \& Rieutord}{Rincon \&
  Rieutord}{2018}]{Rincon2018}
Rincon F.,  Rieutord M.,  2018, \mn@doi [Living Rev. Sol. Phys.]
  {10.1007/s41116-018-0013-5}, 15, 6

\bibitem[\protect\citeauthoryear{{Roudier}, {{\v{S}}vanda}, {Ballot},
  {Malherbe}  \& {Rieutord}}{{Roudier} et~al.}{2018}]{Roudier2018}
{Roudier} T.,  {{\v{S}}vanda} M.,  {Ballot} J.,  {Malherbe} J.~M.,   {Rieutord}
  M.,  2018, \mn@doi [\aap] {10.1051/0004-6361/201732014}, \href
  {https://ui.adsabs.harvard.edu/\#abs/2018A&A...611A..92R} {611, A92}

\bibitem[\protect\citeauthoryear{Rutherford, Dunkerton  \&
  Montgomery}{Rutherford et~al.}{2015}]{Rutherford2015}
Rutherford B.,  Dunkerton T.~J.,   Montgomery M.~T.,  2015, \mn@doi [Q. J. Royal Meteorol. Soc.] {10.1002/qj.2616}, 141, 3344

\bibitem[\protect\citeauthoryear{{Sch\"ussler}}{{Sch\"ussler}}{1984}]{Schussler1984}
{Sch\"ussler} M.,  1984, \aap, \href
  {https://ui.adsabs.harvard.edu/\#abs/1984A&A...140..453S} {140, 453}

\bibitem[\protect\citeauthoryear{{Shadden}, {Lekien}  \& {Marsden}}{{Shadden}
  et~al.}{2005}]{Shadden2005}
{Shadden} S.~C.,  {Lekien} F.,   {Marsden} J.~E.,  2005, \mn@doi [Physica D] {10.1016/j.physd.2005.10.007}, \href
  {https://ui.adsabs.harvard.edu/\#abs/2005PhyD..212..271S} {212, 271}

\bibitem[\protect\citeauthoryear{Silva, Rempel, Gomes, Requerey  \&
  Chian}{Silva et~al.}{2018}]{Silva_2018b}
Silva S. S.~A.,  Rempel E.~L.,  Gomes T. F.~P.,  Requerey I.~S.,   Chian A.
  C.-L.,  2018, \mn@doi [\apj] {10.3847/2041-8213/aad180}, 863, L2

\bibitem[\protect\citeauthoryear{{Stenflo}}{{Stenflo}}{1973}]{Stenflo1973}
{Stenflo} J.~O.,  1973, \mn@doi [\solphys] {10.1007/BF00152728}, \href
  {https://ui.adsabs.harvard.edu/\#abs/1973SoPh...32...41S} {32, 41}

\bibitem[\protect\citeauthoryear{{Stenflo}}{{Stenflo}}{1975}]{Stenflo1975}
{Stenflo} J.~O.,  1975, \mn@doi [\solphys] {10.1007/BF00153287}, \href
  {https://ui.adsabs.harvard.edu/\#abs/1975SoPh...42...79S} {42, 79}

\bibitem[\protect\citeauthoryear{Tanenbaum, Wilcox, Franzier  \&
  Howard}{Tanenbaum et~al.}{1969}]{Tanenbaum1969}
Tanenbaum A.~S.,  Wilcox J.~M.,  Franzier E.~N.,   Howard R.,  1969, \mn@doi
  [\solphys] {10.1007/BF02391655}, 9, 328

\bibitem[\protect\citeauthoryear{{Welsch} \& {Longcope}}{{Welsch} \&
  {Longcope}}{2003}]{Welsch2003}
{Welsch} B.~T.,  {Longcope} D.~W.,  2003, \mn@doi [\apj] {10.1086/368408},
  \href {https://ui.adsabs.harvard.edu/\#abs/2003ApJ...588..620W} {588, 620}

\bibitem[\protect\citeauthoryear{{Yeates}, {Hornig}  \& {Welsch}}{{Yeates}
  et~al.}{2012}]{Yeates2012}
{Yeates} A.~R.,  {Hornig} G.,   {Welsch} B.~T.,  2012, \mn@doi [\aap]
  {10.1051/0004-6361/201118278}, \href
  {https://ui.adsabs.harvard.edu/\#abs/2012A&A...539A...1Y} {539, A1}

\makeatother
\end{thebibliography}

\bsp	
\label{lastpage}
\end{document}